\documentclass[a4paper,fleqn,usenatbib,mn2e,article,scrartcl]{mnras}
\usepackage{newtxtext,newtxmath}
\usepackage[T1]{fontenc}
\usepackage{ae,aecompl}
\usepackage{amsfonts, natbib, txfonts,epsfig,rotating,txfonts,mathtools}

\usepackage{graphicx}	
\usepackage{amsmath}	
\usepackage{amssymb}	
\usepackage{pdflscape,tabularx,footnote}
\makesavenoteenv{tabular}
\makesavenoteenv{table}
\def\Plus{\texttt{+}}
\def\Equ{\texttt{=}}

\title{Narrow vs. Broad line Seyfert 1 galaxies: X-ray, optical and mid-infrared AGN characteristics}

\author[M. Laki\'{c}evi\'{c} et al.]{
Ma\v{s}a Laki\'{c}evi\'{c},$^{1}$\thanks{E-mail: mlakicevic@aob.rs}
Luka \v{C}. Popovi\'{c}$^{1,2}$ and
Jelena Kova\v{c}evi\'{c}-Doj\v{c}inovi\'{c}$^{1,2}$ 
\\
$^{1}$Astronomska opservatorija Beograd; Volgina 7, 11060 Beograd, Serbia\\
$^{2}$Isaac Newton Institute of Chile, Yugoslavia Branch\\
}

\date{Accepted: 14 May 2018}

\pubyear{2018}

\begin{document}
\label{firstpage}
\pagerange{\pageref{firstpage}--\pageref{lastpage}}
\maketitle

\begin{abstract}
We investigated narrow line Seyfert 1 galaxies (NLS1s) at optical, mid-infrared (MIR) and X-ray wavelengths, comparing them to the broad line  active galactic nuclei (BLAGNs). We found that black hole mass, coronal line luminosities, X-ray hardness ratio and X-ray, optical and MIR luminosities are higher for the BLAGNs than for NLS1s, while policyclic aromatic hydrocarbon (PAH) contribution and the accretion rates are higher for the NLS1s. Furthermore, we found some trends among spectral parameters that NLS1s have and BLAGNs do not have. The evolution of FWHM(H$\beta$) with the luminosities of MIR and coronal lines, continuum luminosities, PAH contribution, H$\beta$ broad line luminosity, FWHM[O\,III] and EW(H$\beta$NLR), are important trends found for NLS1s. That may contribute to the insight that NLS1s are developing AGNs, growing their black holes, while their luminosities and FWHM(H$\beta$) consequently grow, and that BLAGNs are mature, larger objects of slower and/or different evolution. Black hole mass is related to PAH contribution only for NLS1s, which may suggest that PAHs are more efficiently destroyed in NLS1s.
\end{abstract}


\section{Introduction} \label{sec:intro}
A specific group of Type 1 active galactic nuclei (AGNs), narrow line Seyfert galaxies (NLS1s) have the broad emission lines in the optical part of spectra, although the broad lines are narrower than in the spectra of other Type 1 objects, so called Broad line Seyfert 1 galaxies -- BLS1 \citep{Zhou06,Tarchi11}. Not much is known about the nature of these objects: their ambiguous characteristics challenge the Unified Model of AGN. NLS1s are less often radio-loud \cite[only $\sim$7\%;][]{Komossa06}, have relatively strong optical Fe\,II emission, weak [O\,III] emission, denser broad line region (BLR) clouds, large CIV blueshifts, and at X-rays -- NLS1s have strong soft excess emission below 1 keV and a rapid flux variability, compared to BLS1s \citep{Zhou06,Nikolajuk09,Tarchi11,Shapovalova12,Rakshit17a}. NLS1s have a lower optical variability \citep{Rakshit17b} and possibly lower average inclination than the BLS1s \citep{Liu16,Rakshit17a}. NLS1s are fainter than BLS1s in the UV and at hard X-ray \citep{Grupe10}. \citet{Jarvela15} showed that NLS1 sources with higher black hole masses (M$_{\rm BH}$s) are more likely to be able to launch a powerful relativistic jet and be detected at $\gamma$-rays.

It is found that the accretion rate in units of the Eddington accretion rate for NLS1s is larger than for the broad line AGNs (BLAGNs) \cite[and references therein]{Bian03}. That is consistent with the finding that the NLS1s are hosted mostly in barred Seyfert galaxies, as 65\% of the NLS1s have bars, while only 25\% of BLS1s have bars \citep{Crenshaw03}. These stellar bars could be an efficient way of transporting large amounts of gas and dust to the inner region \citep{Deo06}. \citet{Deo06} found that 80\% of the NLS1s have large nuclear dust spirals, while 32\% BLS1s have these structures.

The usual estimation is that NLS1s have lower M$_{\rm BH}$ than other Seyfert 1 (S1) galaxies \citep{Mathur00,Wang01,Deo06,Jin12a}. It is believed that NLS1s are regular Seyfert galaxies at an early stage of evolution \citep{Mathur00,Mathur01} and that their black holes may still be growing \cite[see][and references therein]{Jin12a}. However, some other studies show that NLS1s have similar M$_{\rm BH}$ and Eddington ratio to BLS1s, but difference between them is in the geometry \citep{Baldi16,Liu16}. Additionally, \citet{Rodriguez97} noticed NLS1s with the high-ionization UV permitted lines, Ly$\alpha$, C\,IV $\lambda$1550, and He\,II $\lambda$1640 with full width at half maximum (FWHM) of broad components $\sim$ 5000 km s$^{-1}$. Therefore, NLS1s do contain gas that is moving at velocities comparable to those found in S1 galaxies. 

The efforts have been made to understand the origin of the broad lines in NLS1s, as e.g. MRK\,493 has the weak broad component detected in the H$\alpha$, H$\beta$, and He $\lambda$4686, that is probably from the BLR, but also may be produced by violent starburst \citep{Popovic09}. Also, they found that the source of ionization of the narrow Balmer and [O\,III] lines seems not to be the AGN, but the star-forming processes. As well, in radio-quiet NLS1s, the radio and infrared emission is more likely generated via the star formation processes, than by the central source \citep{Jarvela15}. Additionally, in a group of optically detected Type 1 AGNs, \citet{Popovic11} found that starburst dominated galaxies have several correlations different from the rest of objects, such as the correlation between FWHM of broad emission line H$\beta$ -- FWHM(H$\beta$) and the optical AGN continuum luminosity at 5100$\AA$ (L5100), which suggests the relation between BLR and the ionization source.

When we summarize all these facts, there are several scenarios or reasons why NLS1s differ from BLAGNs: 1) They have more starbursts and therefore some correlations are different \citep{Popovic11}; 2) They may be young S1 galaxies in the early stage of evolution, where the black hole is still growing \citep{Grupe04,Jin12a}; 3) Some estimations show a lower M$_{\rm BH}$ in NLS1s than in BLS1s \citep{Wang01}; 4) They have more bars and dust spirals \citep{Deo06} and higher accretion rate \citep{Bian03}; 5) NLS1s may be ordinary Seyfert galaxies with a different geometry \citep{Baldi16,Liu16}; 6) There may be some other reason why they are different.

Mid infrared (MIR) forbidden emission lines, such as [Ne\,V], [Ne\,III], [O\,IV], and [SIV], originate from the photoionizations and excitations of ions in the low density interstellar medium of the host galaxy, narrow line region (NLR) \citep{Dasyra11,RodriguezArdila11,Hill14}. The fluxes of MIR lines [Ne\,V], [O\,IV], [Ne\,III], and [SIV] correlate among each other and with L5100 and M$_{\rm BH}$ \citep{Dasyra11}. [O\,IV] at 25.89 $\mu$m, and [Ne\,V] at 14.32 and 24.32 $\mu$m belong to the coronal lines, as they have a high ionization potential that can hardly be excited by the other radiation than the AGN, 54 and 97 eV, respectively. [Ne\,III] at 15.55 $\mu$m (41 eV) is not coronal line.

Seyfert galaxies are lower luminosity AGNs with absolute magnitude M$\rm _{B}$>-21.5+logh$_{0}$, where h$_{0}$ is the Hubble constant in units of 100 km s$^{-1}$ Mpc$^{-1}$, while quasars are the most luminous AGNs, with M$\rm _{B}$<-21.5+logh$_{0}$ \citep{Schmidt83}. In this study we use the term broad line AGNs (BLAGNs) and there we include both BLS1 and all quasars with broad lines (FWHM(H$\beta$)$\ge$ 2000 km s$^{-1}$; see Sec.~\ref{sec:sam}).

Here we compare some MIR, optical and X-ray spectral characteristics of BLS1s and BLAGNs, and their correlations, in order to explore their structures and activities. We find some differences that may help in understanding the physics of these objects.

This paper is organized as following: in Sec.~\ref{sec:sample} we describe the sample and the method of the analysis, in Sec.~\ref{sec:results} we present the results, in Sec.~\ref{sec:discuss} we discuss the results, and in Sec.~\ref{sec:conclussion} we list our conclusions.

\section{The sample and method of analysis} \label{sec:sample}
\subsection{The sample} \label{sec:sam}
In the first part of this study, Sec.~\ref{sec:prvo} and \ref{sec:prvoi}, we analysed the differences between NLS1s and BLS1s, using the AGN sample from \citet{Sani10}, consisting from 60 NLS1s and 55 BLS1s, from which only 49 NLS1s and 54 BLS1s were used, as they gave satisfactory fits using {\sl deblendIRS} routine (see Sec.~\ref{sec:met}). These objects were carefully chosen as S1 galaxies, as non-radio loud quasars classified as S1n, S1, S1.0 or S1.2, with redshift $\le$0.2, that are observed with InfraRed Spectrograph -- IRS \citep{Houck04} from the Spitzer Space telescope. Their criterion for NLS1s was that FWHM(H$\beta$)$\le$2000 km s$^{-1}$ \citep{Veron01}, and that the ratio of total [OIII]$\lambda$5007 to total H$\beta$ is <3 \citep{VC06}. In this work we adopted FWHM(H$\beta$)$\le$2200 km s$^{-1}$, as it is given in \citet{Rakshit17a}, and that total [OIII]$\lambda$5007 to total H$\beta$ <3.

In the second part of this study, Sec.~\ref{sec:drugo}, \ref{sec:novi} and \ref{sec:next}, we compared the characteristics of 64 NLS1 and 99 BLAGN objects from \citet{Sani10}, \citet{Lakicevic17}, \citet{Jin12b}, \citet{Rakshit17a}, \citet{Nikolajuk09}, \citet{McLure02} and \citet{Whalen06}, that were observed by the IRS on Spitzer, and some optical data exist in the literature mentioned above. For the objects from \citet{Lakicevic17} the criterion for NLS1s was that FWHM(H$\beta$)$\le$ 2200 km s$^{-1}$ \citep{Rakshit17a}, and that the flux ratio of total [OIII]$\lambda$5007 to total H$\beta$ is <3 \citep{Osterbrock85}. The fluxes of [OIII]$\lambda$5007 and total H$\beta$, and their ratios, for 7 objects from \citet{Lakicevic17}, that were not previously classified as NLS1s, are given in the Appendix~\ref{appen}. We concluded that these objects are NLS1s. The rest of the objects are classified as it is given in corresponding literature (Table~\ref{tab1sani10laki}). 

\subsection{The Method} \label{sec:met}
82 objects from the sample were already analysed in \citet{Lakicevic17} and therefore we took the optical and MIR parameters from that work. For the rest of the data, we calculated MIR parameters (see below), and adopted the optical data (FWHM(H$\beta$) and L5100) from the literature (Table~\ref{tab1sani10laki}).

The low-resolution IRS spectra \citep{Lebouteiller11}, fully reduced and calibrated, were downloaded from the Combined Atlas of Sources with Spitzer IRS Spectra -- CASSIS\footnote{\url{http://cassis.sirtf.com/}} database. The spectra were fitted in {\sl deblendIRS} code \citep{Hernan15}. This routine, written in {\sc IDL}, performs the decomposition of spectra to AGN, PAH and stellar part. More about this analysis can be found in \citet{Lakicevic17}. The products of the fitting were MIR parameters: fractional contributions of AGN, PAH and stellar components to the integrated 5--15$\mu$m luminosity, named RAGN, RPAH and RSTR, respectively (RAGN + RPAH+ RSTR=1), spectral index of the AGN component -- $\alpha$, the silicate strength -- $\rm S_{SIL}$\footnote{Logarithm of the ratio of the flux densities of the spectrum and the underlying continuum, at the wavelength of the peak.}, $\chi^2$, the coefficient of variation of the rms error ($\rm CV_{RMSE}$) and monochromatic luminosity of the source at 6 $\mu$m, hereafter L6 \citep{Hernan15,Lakicevic17}. The resulting MIR parameters are given in Table~\ref{tab1sani10laki}.

\begin{table*}
\caption{The MIR and optical data for the NLS1s and BLAGNs, respectively. Column (1) -- source name, (2) -- redshift, (3) -- FWHM(H$\beta$) = FWHM of broad H$\beta$ component, (4) -- logarithm of AGN luminosity at 5100$\AA$, multiplied with 5100, (5) and (6) -- fractional contributions of AGN and PAH components to the integrated MIR luminosity, respectively, (7) -- silicate strength, (8) -- spectral index, (9) -- monochromatic luminosity of the source at 6$\mu$m, (10)-- Luminosity of H$\beta$ broad line (ILR+VBLR -- see the text) for the sources from \citet{Lakicevic17}, (11) logarithm of black hole mass, (12) -- a reference for the optical data: Sani10=\citet{Sani10}, Laki17=\citet{Lakicevic17}, VC06=\citet{VC06}, Jin12=\citet{Jin12b}, Rak17=\citet{Rakshit17a}, Nik09=\citet{Nikolajuk09}, MJ02=\citet{McLure02} and W06=\citet{Whalen06}.  A full version is available in the electronical form. \label{tab1sani10laki}}
\centering{
\begin{tabular}{|l|c|c|c|c|c|c|c|c|c|c|c|}
\hline\hline
NAME     &    z     &   FWHM(H$\beta$)  &Log$\lambda$L5100    &RAGN& RPAH&   SSIL&$\alpha$& L6         &    LH$\beta$b     &log(M$_{\rm BH}$) &ref\_opt\\
         &          &   km s$^{-1}$     &erg s$^{-1}$&    &     &       &        &erg s$^{-1}$&erg s$^{-1}$&[M$_{\odot}$] & \\
(1) &(2)&(3)&(4)&(5) &(6)&(7)&(8)&(9) &(10)&(11)&(12)\\ \hline
\hline  
Ark564              &0.025&    865 &     44.12&      0.958& 0.033   &0.1    &-1.443   &4.578E43 &--&6.67&Sani10\\
IRAS13224-3809      &0.067&    650 &     44.81&      0.674& 0.201   &-0.077 &-2.043   &1.115E44 &--&6.87&Sani10\\
1H0707-495          &0.041&    1000&     43.48&      0.907& 0.014   &0.254  &-1.301   &2.237E43 &--&6.38&Sani10\\\hline
Fairall9            &0.047&    6690&     43.92&      0.875& 0.006   &0.122  &-1.237   &3.951E44 &--& 8.32&Sani10\\
IRAS13342+3932      &0.179&    7090&     44.42&      0.760& 0.204   &0.243  &-2.094   &8.296E44 &--&8.69&Sani10\\
FBS0732+396         &0.118&    2370&     44.31&      0.800& 0.026   &0.157  &-1.428   &4.672E44 &--&7.67&Sani10\\ \hline
\end{tabular}}
\\
\smallskip
\end{table*}

The luminosities of the H$\beta$ broad line (the sum of intermediate line region -- ILR and very broad line region -- VBLR components) for the sources from \citet{Lakicevic17} are the only optical parameters calculated for this work (Table~\ref{tab1sani10laki}). Their optical data are available in the SDSS Data Release 12 (DR12) \cite[see more in][]{Lakicevic17}.

We used X-ray data from the 3XMM-DR5 version of XMM-Newton serendipitous survey \citep{Rosen16}, that provides measured fluxes and hardness ratios (HRs). HR is defined as HR1,2=(F2-F1)/(F2+F1), where F2 is the flux of the harder band (higher energy) than F1. The indices 1, 2, 3, 4 and 5 mark the narrow bands: 1 = 0.2 -- 0.5 keV, 2 = 0.5 -- 1.0 keV, 3 = 1.0 -- 2.0 keV, 4 = 2.0 -- 4.5 keV and 5 = 4.5 -- 12.0 keV. In such a way the hardest sources would have HR of 1, while the softest would have HR of -1. We found HRs and the luminosities at 0.2-12 keV for the 33 NLS1 and 33 BLS1 objects (Table~\ref{tab3}).

\begin{table*}
\caption{Hardness ratios (HRs) and the luminosities at 0.2-12 keV for NLS1s and BLS1s. The indices 1-5 indicate the narrow bands among which the HRs were calculated (see the text). The first 33 are NLS1s, while the rest 33 objects are BLS1s. A full version is available in the electronical form. \label{tab3}}
\centering{
\begin{tabular}{|l|c|c|c|c|c|c|c|}
\hline\hline
NAME                &z   &     IAUNAME(JHHMMSS.s$+$DDMMSS) & HR1\_1,2 &HR2\_2,3 &HR3\_3,4 &HR4\_4,5& L$_{\rm 0.2-12 keV}$\\
(1) &(2)&(3)&(4)&(5) &(6)&(7)&(8)\\ \hline
\hline
  Ark564             & 0.0247&   J224239.3+294331&           0.212  & -0.124&  -0.572&  -0.686 &1.08E44\\
  IRAS13224-3809     & 0.0667&   J132519.3-382452&           0.0060 & -0.565&  -0.603&  -0.567 &3.52E43 \\
  1H0707-495         & 0.0411&   J070841.4-493306&           0.135  & -0.563&  -0.638&  -0.627 &2.20E43\\ \hline \hline
 Fairall9            &0.047   &J012345.7-584820&              0.238&   0.088&   -0.357&  -0.487 &1.43E44\\ 
 J131305.68-021039.2 &0.0837  &J131305.7-021039&              0.224&   -0.036&  -0.405&  -0.332 &8.35E43\\
 PG0052+251          &0.155   &J005452.1+252539&              0.242&   0.073 &  -0.387&  -0.564  &9.1E44\\ \hline
\hline
\end{tabular}}
\\
\smallskip
\end{table*}

The Eddington ratio (measure of the accretion rate in Eddington units) from the optical data is found as R$_{\rm Edd}^{\rm opt}$ = L$_{\rm BOL}$ L$_{\rm Edd}^{-1}$, where bolometric luminosity is given as L$_{\rm BOL}\sim$ 9$\cdot$L5100, and Eddington luminosity as L$_{\rm Edd}$ = 1.26$\times$10$^{46}$erg s$^{-1}$M$_{\rm BH}$ [10$^{8}$M$\odot$] \citep{Wu04}.

\begin{figure}  
\centering
\rotatebox{0}{
\includegraphics[width=91mm]{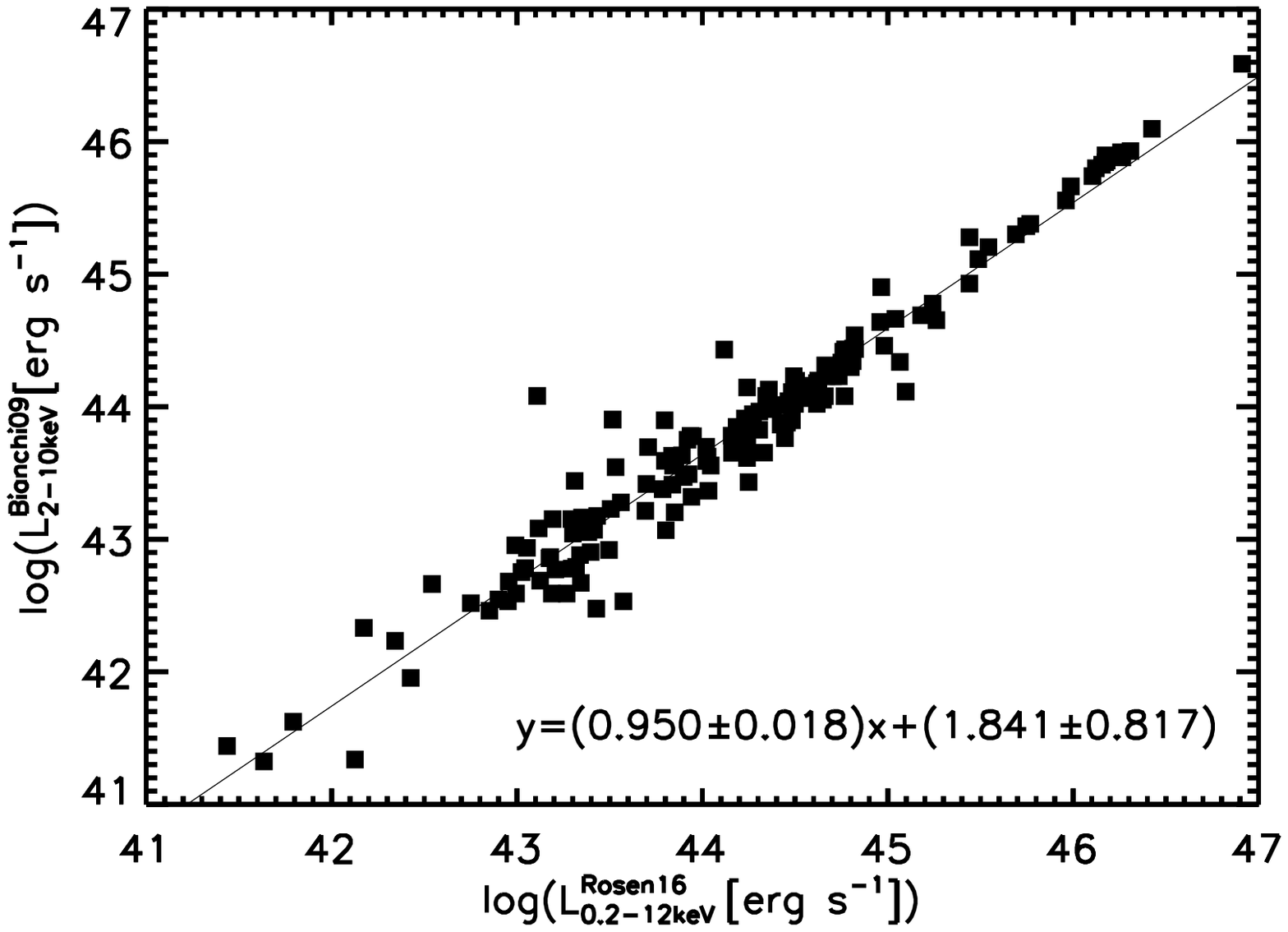}}
\caption{The relationship between 0.2-12 keV \citep{Rosen16} and 2-10 keV luminosities \citep{Bianchi09}.   \label{fig:conv}}
\end{figure}

Similarly as above, Eddington ratio from the X-ray data is found as R$_{\rm Edd}^{\rm X}$ = L$_{\rm BOL}$L$_{\rm Edd}^{-1}$, where L$_{\rm BOL}\sim$25$\cdot$L$_{\rm 2-10 keV}^{\rm X}$ \citep{Georgakakis17}, and L$_{\rm Edd}$ is the same as above. Here we needed the luminosity at 2-10 keV band, but we only had the luminosities from 0.2-12 keV band from \citet{Rosen16} (see Section~\ref{sec:sam}). Therefore, we found the conversion from 0.2-12 keV to 2-10 keV luminosity, using the catalogue \citet{Bianchi09} (where they gave 2-10 keV luminosities) and catalogue \citet{Rosen16}, where they have luminosities at 0.2-12 keV band. That conversion is L$_{\rm 2-10 keV}^{\rm X}$=(0.950$\pm$0.018)L$_{\rm 0.2-12 keV}^{\rm X}$ + (1.841$\pm$0.817) (see Fig.~\ref{fig:conv}). 

\begin{figure}  
\centering
\rotatebox{90}{
\includegraphics[width=67mm]{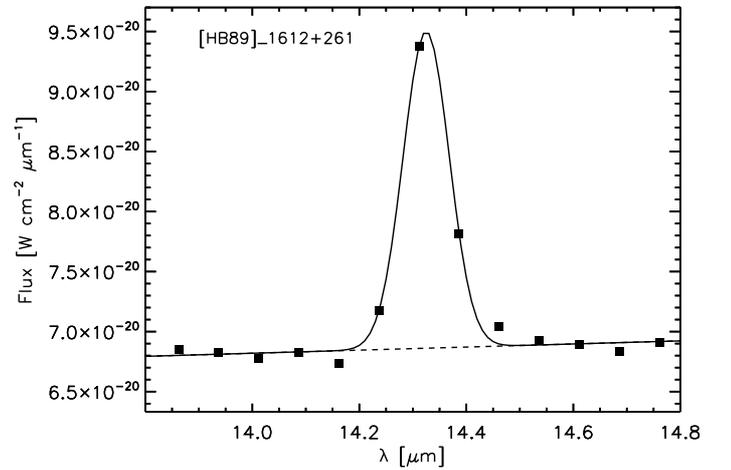}}
\caption{The fit of the coronal line [Ne\,V] at 14.3 $\mu$m with a Gaussian function.   \label{fig:fit}}
\end{figure}

The MIR lines were fitted with a simple Gaussian (see Fig.~\ref{fig:fit}), and their fluxes and luminosities were found. We adopted the fitting results that show good fits (signal to noise>3). The luminosities of MIR lines [Ne\,V] at 14.32 $\mu$m, [O\,IV]\footnote{Since in the IRS low resolution it is not possible to distinguish [O\,IV] 25.89 $\mu$m line coming from the AGN and/or intense star formation \citep{Dixon11} from the [Fe\,II] 25.99 $\mu$m line -- arising from the star formation, we assume that we detect and measure [O\,IV] 25.89 $\mu$m line \citep{Melendez08}.} at 25.89 $\mu$m, and [Ne\,III] line at 15.55 $\mu$m are given in Table~\ref{tab:lum}. We chose these lines because [Ne\,V] and [O\,IV] are coronal lines and because the lines at these three wavelengths have most detections. 32 from 64 NLS1s and 55 from 99 BLAGNs are detected at one or more MIR lines. 

\begin{table}
\caption{The luminosities of MIR lines for both NLS1 and BLAGN samples: [O\,IV] at 25.89 $\mu$m, [Ne\,V] at 14.32 $\mu$m, and [Ne\,III] at 15.55 $\mu$m. A full version is available in the electronical form. \label{tab:lum}}
\centering{
\begin{tabular}{|l|c|c|c|c|c|c|}
\hline\hline
NAME     &   L25$\mu$m    &     L14$\mu$m &     L15$\mu$m \\
         &   [erg s$^{-1}$] &   [erg s$^{-1}]$  &  [erg s$^{-1}$]   \\
\hline  
 Fairall9               & 3.51E41 &--     & 2.21E41   \\
 IRAS13342+3932         & 8.04E42 &3.28E42& 3.42E42   \\
 FBS0732+396            & 2.51E42 &4.99E41& 9.72E41   \\
 J131305.68-021039.2    & 2.16E41 &--     & 1.29E41   \\ \hline
\end{tabular}}
\\
\smallskip
\end{table}

In Sec.~\ref{sec:results} we used Mann-Whitney U-test (Wilcoxon Rank-Sum Test) in order to compare the parameters among NLS1 and BLS1 objects. This test is used to reveal whether the two independent samples are selected from the same population, by comparing the medians of these two samples. It does not require the assumption of normal distributions, like the T-test. Null hypothesis (H0) for this test is that the two populations have the same distribution. The main result of this test is the P value. If the computed P is greater than the chosen significance level (we adopted 0.05), H0 is accepted, else H0 is rejected.  

\section{Results} \label{sec:results}
\begin{figure} 
\centering
\rotatebox{90}{
\includegraphics[width=65mm]{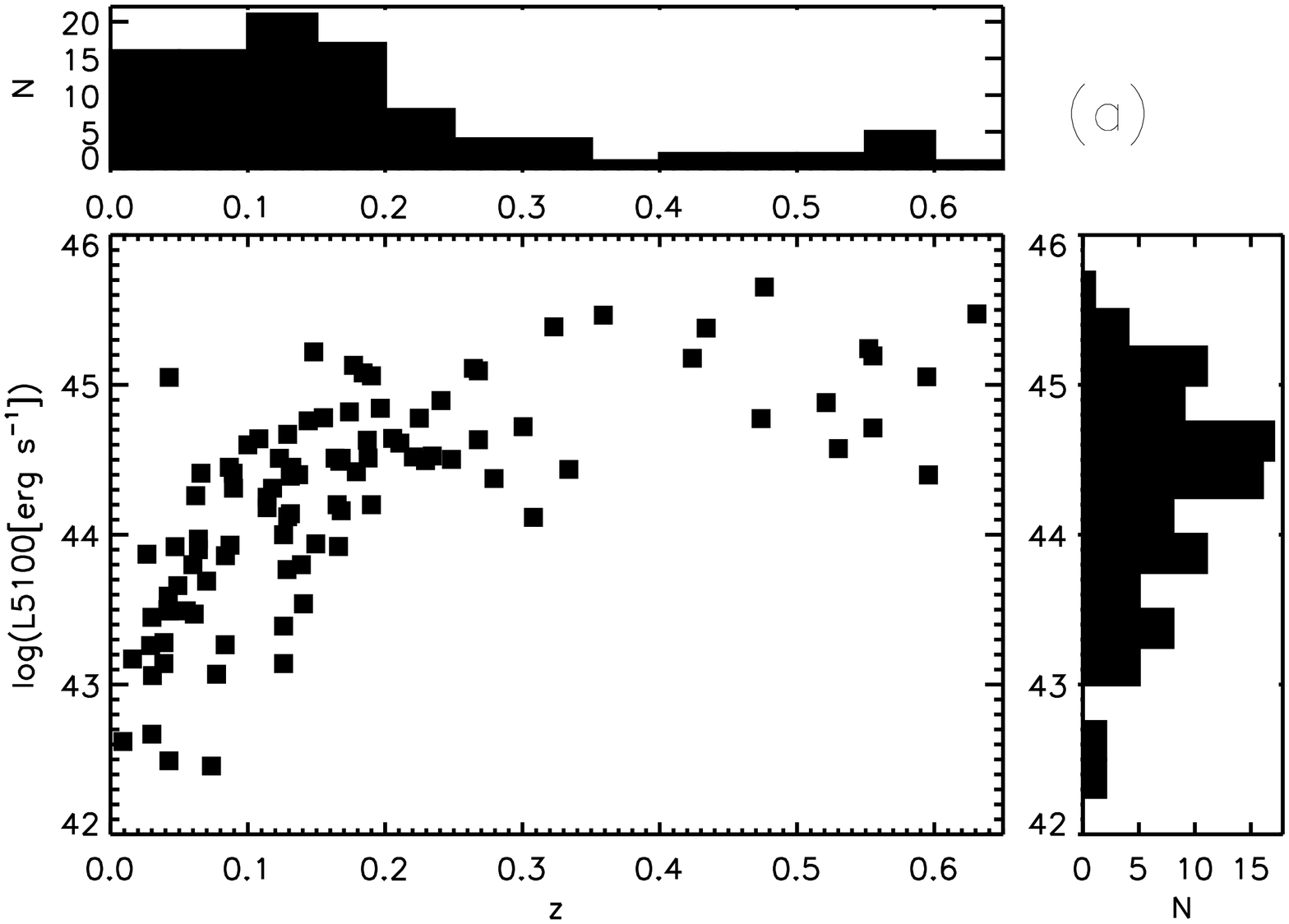}}
\rotatebox{90}{
\includegraphics[width=65mm]{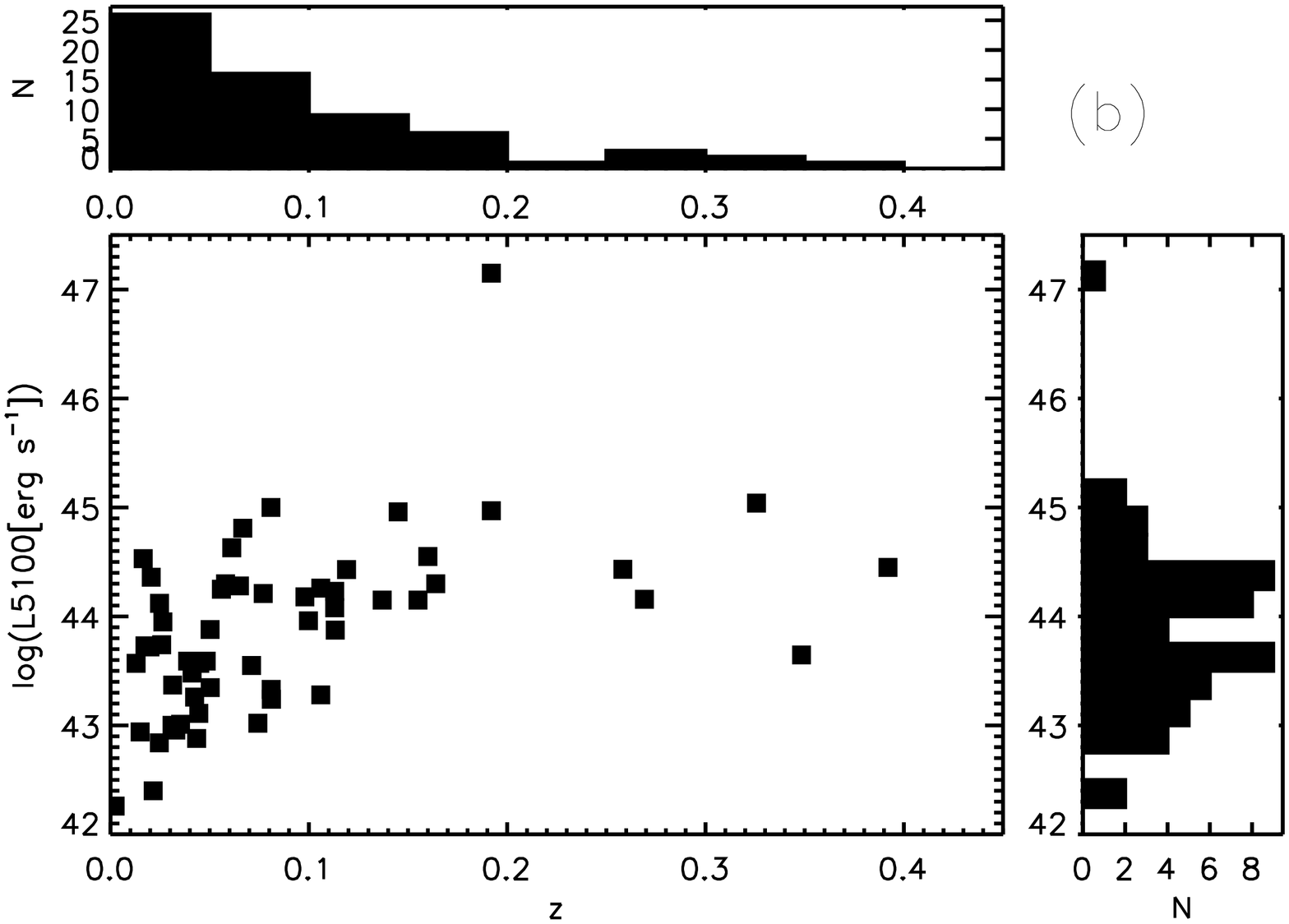}}
\caption{The distributions of the redshifts and AGN luminosities, L5100 for the samples of: (a) BLAGNs and (b) NLS1s. \label{fig:z}}
\end{figure}

In the Fig.~\ref{fig:z} we present the distributions of redshifts and L5100 for the total sample of (a) BLAGNs and (b) NLS1s. As expected, redshift is correlated with L5100, somewhat stronger for BLAGNs: R=0.77, P<0.00001, than for NLS1s: R=0.56, P<0.00001. 

We performed Mann-Whitney U-test in order to compare the redshifts between NLS1 and BLAGN objects. The resulting P value of this test is P<0.00001, which means that the redshifts are significantly different. NLS1s have somewhat lower redshifts: mean(z$\rm_{BLAGNs}$)=0.184, median(z$\rm_{BLAGNs}$)= 0.140, mean(z$\rm_{NLS1s}$)=0.095, median(z$\rm_{NLS1s}$)=0.0667. We do not consider this difference particularly important for the characteristics of these objects -- see more in Sec.~\ref{sec:prvoi}. If we exclude one NLS1 with the highest L5100 (>47), BLAGNs are distributed over the larger interval than NLS1s. Mann-Whitney test for L5100 between NLS1s and BLAGNs shows P=0.0005, which means that L5100 is significantly higher for BLAGNs: mean(L5100$\rm_{BLAGNs}$)=44.26, median(L5100$\rm_{BLAGNs}$)= 44.41, than for NLS1s: mean(L5100$\rm_{NLS1s}$)= 43.89, median(L5100$\rm_{NLS1s}$)=43.91.

\subsection{Comparison of NLS1s and BLS1s} \label{sec:prvo}
It is known that NLS1s have more often PAH molecules than the BLS1s: PAHs are found in 70\% NLS1, and 45\% BLS1 objects \citep{Sani10}. Also, NLS1s have higher accretion rate (see Section~\ref{sec:intro}). Mann-Whitney U-test is performed in order to compare all spectroscopic parameters between NLS1s and BLS1s. As one can see from Table~\ref{tab2}, all parameters (except RPAH and the optically derived accretion rate, R$_{\rm Edd}^{\rm opt}$) are higher for BLS1s than for NLS1s. The luminosities are often higher one order of magnitude. The optical and MIR luminosities, AGN contribution, redshift, and M$_{\rm BH}$ are all higher for the BLS1s than for NLS1s. The parameters from the Table~\ref{tab1sani10laki} that are not given in Table~\ref{tab2} do not show any significant differences according to this test, or they are not available for enough objects.

\begin{table*}
\caption{Man-Whitney U-test for significant differences between optical and MIR parameters of NLS1 and BLS1 objects: the parameter names are given in the column (1), P values are given in the column (2). Mean values of the parameters for the NLS1s and BLS1s are given in the columns (3) and (4), while their medians are given in the columns (5) and (6).  \label{tab2}}
\centering{
\begin{tabular}{|c|c|c|c|c|c|}
\hline\hline
Parameter &P value& Mean NLS1s& Mean BLS1s& Median NLS1s& Median BLS1s\\
(1) &(2)&(3)&(4)&(5) &(6)\\
\hline
z (redshift)   &     0.00158  &0.076                  & 0.105                 &0.058                &0.114\\ \hline 
L5100          &     0.03078  &43.89                  &44.137                 &43.95                &44.2\\ \hline 
L6             &      0.00124 &2.83E+44               &6.00E+44               &6.58E+043            &2.67E+044\\ \hline  
RPAH           &      0.0003  &0.101                  &0.032                  &0.0525               &0.017 \\ \hline  
RAGN           &      0.034   &0.797                  &0.856                  &0.792                &0.850\\ \hline  
M$_{\rm BH}$   &      <0.00001& 6.85                  &8.06                   &6.86                 &8.11\\ \hline  
R$_{\rm Edd}^{\rm opt}$  & <0.00001                  &1.13                  &0.126                 &0.783  &0.097 \\ \hline  
L[Ne\,V]14.32$\mu$m&  0.013   &6.78E+40               &4.29E+41             & 3.54E+40              &2.05E+41 \\ \hline 
L[O\,IV]25.89$\mu$m&  0.00452 &2.31E+41               &9.94E+41             &1.25E+41            &3.71E+41
\\ \hline
\hline
\end{tabular}}
\\
\smallskip
\end{table*}

\subsubsection{NLS1s and BLS1s with X-ray data} \label{NLS1sBLS1s}
All four HRs and luminosities at 0.2-12 keV band, L$_{\rm 0.2-12 keV}$, were compared for the NLS1 and BLS1 samples, and significant differences were found in 3 HRs and luminosity L$_{\rm 0.2-12 keV}$ (Table~\ref{tab5}). NLS1 objects have lower HRs and luminosities than BLS1s. That indicates that NLS1s are softer and less bright sources at X-rays. The accretion rate R$_{\rm Edd}^{\rm X}$, derived from the X-ray data (see Section~\ref{sec:sample}) is higher for NLS1s than for BLS1s, as shown in the literature (see Table~\ref{tab5}, and Section~\ref{sec:intro}).

\begin{table*}
\caption{Man-Whitney U-test for the differences between NLS1 and BLS1 objects for X-ray data: hardness ratios (HRs), luminosity at 0.2-12 keV and the accretion rate derived from the X-ray data. The parameter names are given in the column (1), P values are given in the column (2). Mean values of the parameters for the NLS1s and BLS1s are given in the columns (3) and (4), while their medians are given in the columns (5) and (6).  \label{tab5}}
\centering{
\begin{tabular}{|c|c|c|c|c|c|}
\hline\hline
        &P value& Mean NLS1s& Mean BLS1s& Median NLS1s& Median BLS1s\\
(1) &(2)&(3)&(4)&(5)&(6) \\
\hline
HR 2,3  &  0.00014 &  -0.212& 0.0056 & -0.204& -0.021\\ \hline
HR 3,4  &    0.00016&  -0.401 & -0.291 & -0.494& -0.357\\ \hline
HR 4,5  &  0.0007 &  -0.490 & -0.433 & -0.556& -0.483\\ \hline
L$_{\rm 0.2-12 keV}$ & 0.00068 & 1.12$\times$10$^{44}$ & 2.28$\times$10$^{44}$ &2.21$\times$10$^{43}$ &1.37$\times$10$^{44}$ \\ \hline
R$_{\rm Edd}^{\rm X}$ & 0.00014 &1.094 & 0.145 & 0.432 &0.098 \\ \hline
\hline
\end{tabular}}
\\
\smallskip
\end{table*}

\subsection{Comparison NLS1s and BLAGNs} \label{sec:drugo}
The magnitudes of the spectroscopic parameters for NLS1s and BLS1s were compared in Sec.~\ref{sec:prvo}. Since quasars (they belong in BLAGN) are brighter and more massive objects than S1s, here we explore only the correlations of NLS1s vs. BLAGNs. Most of the conclusions from Sec.~\ref{sec:prvo} about BLS1s could also apply for the BLAGNs.    

\subsubsection{Correlations of optical, X-ray and MIR parameters of NLS1s and BLAGNs}
\begin{figure*}  
\centering
\rotatebox{90}{
\includegraphics[width=65mm]{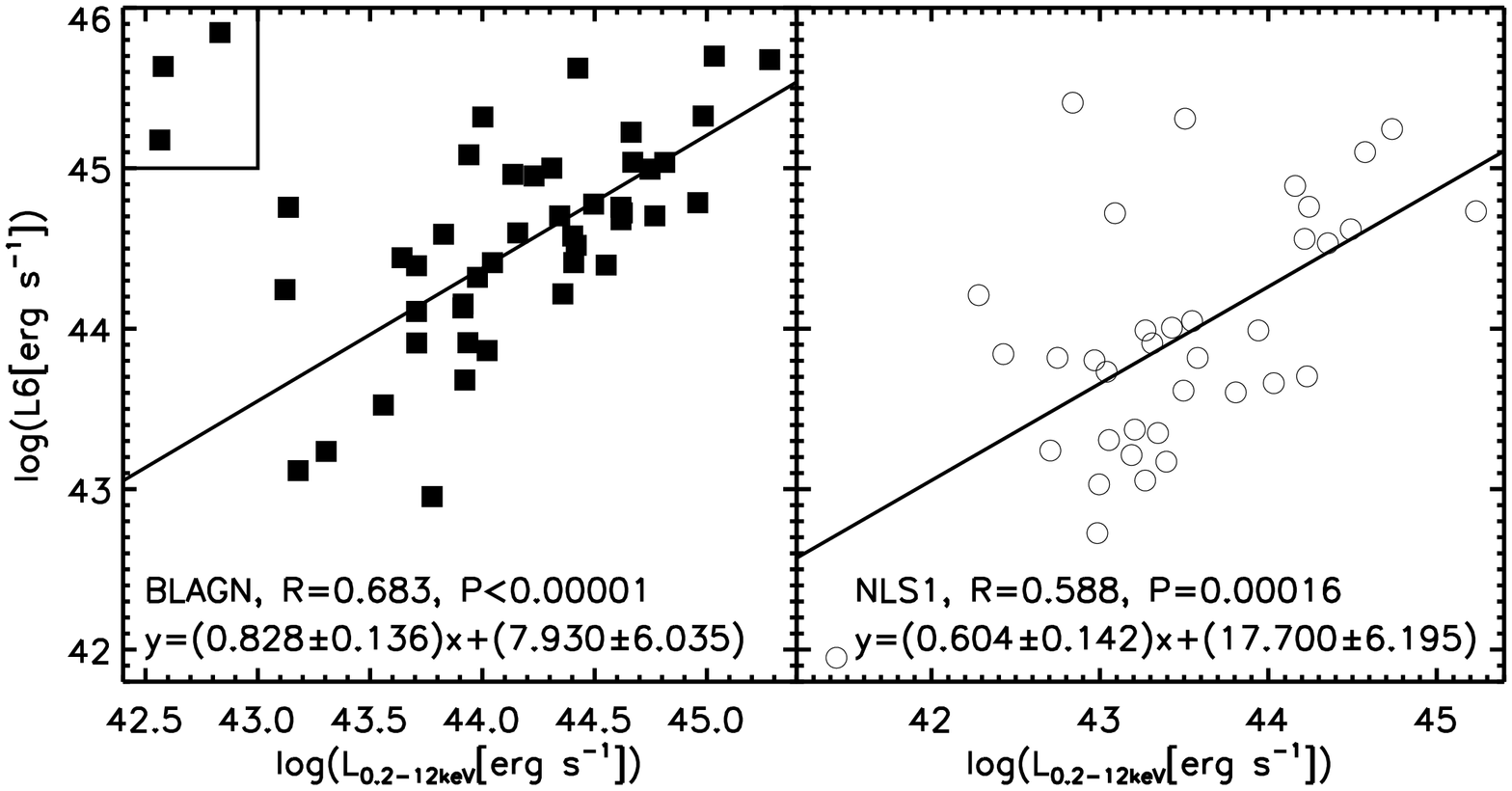}}
\rotatebox{90}{
\includegraphics[width=65mm]{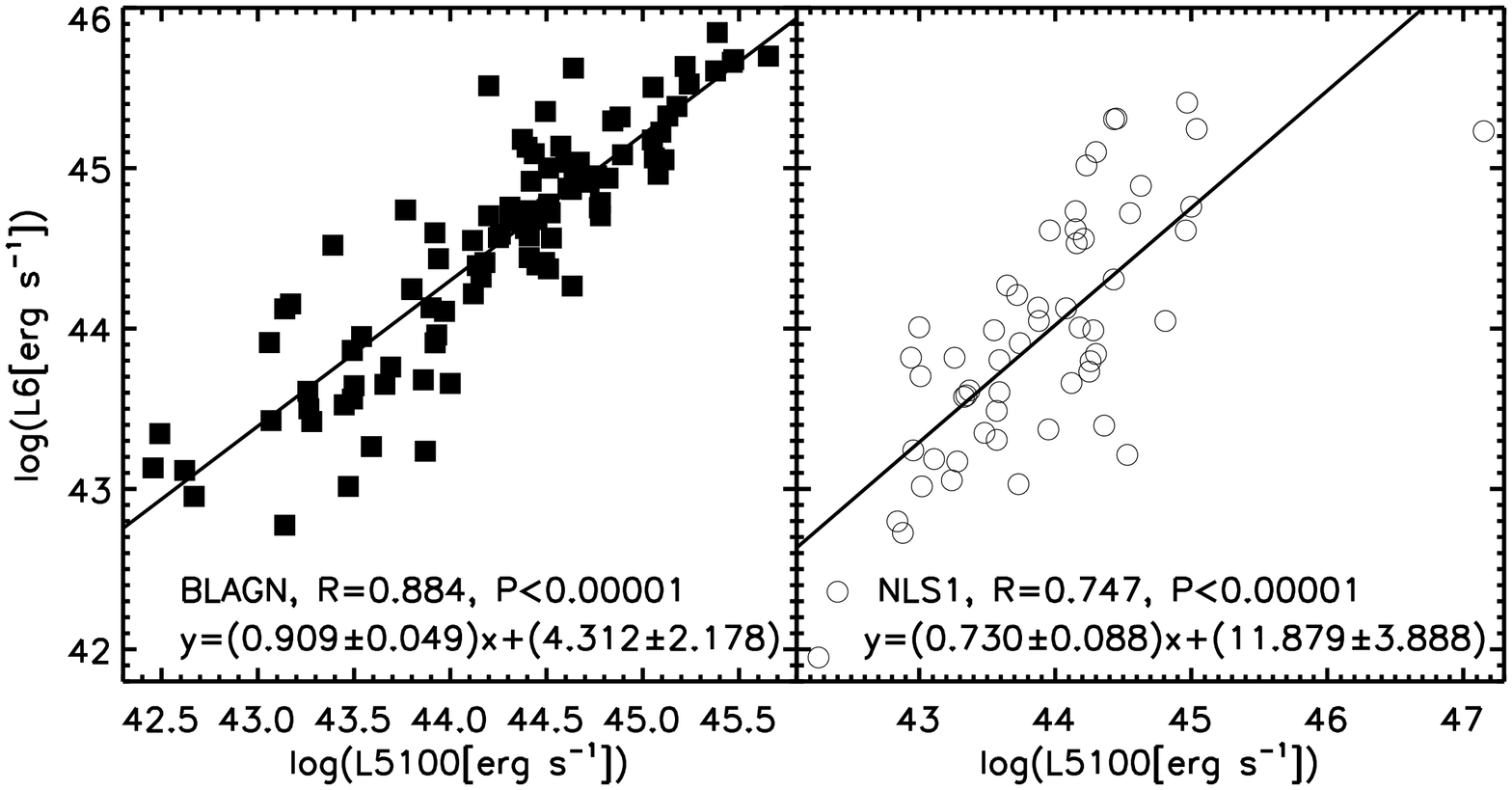}}
\rotatebox{90}{
\includegraphics[width=65mm]{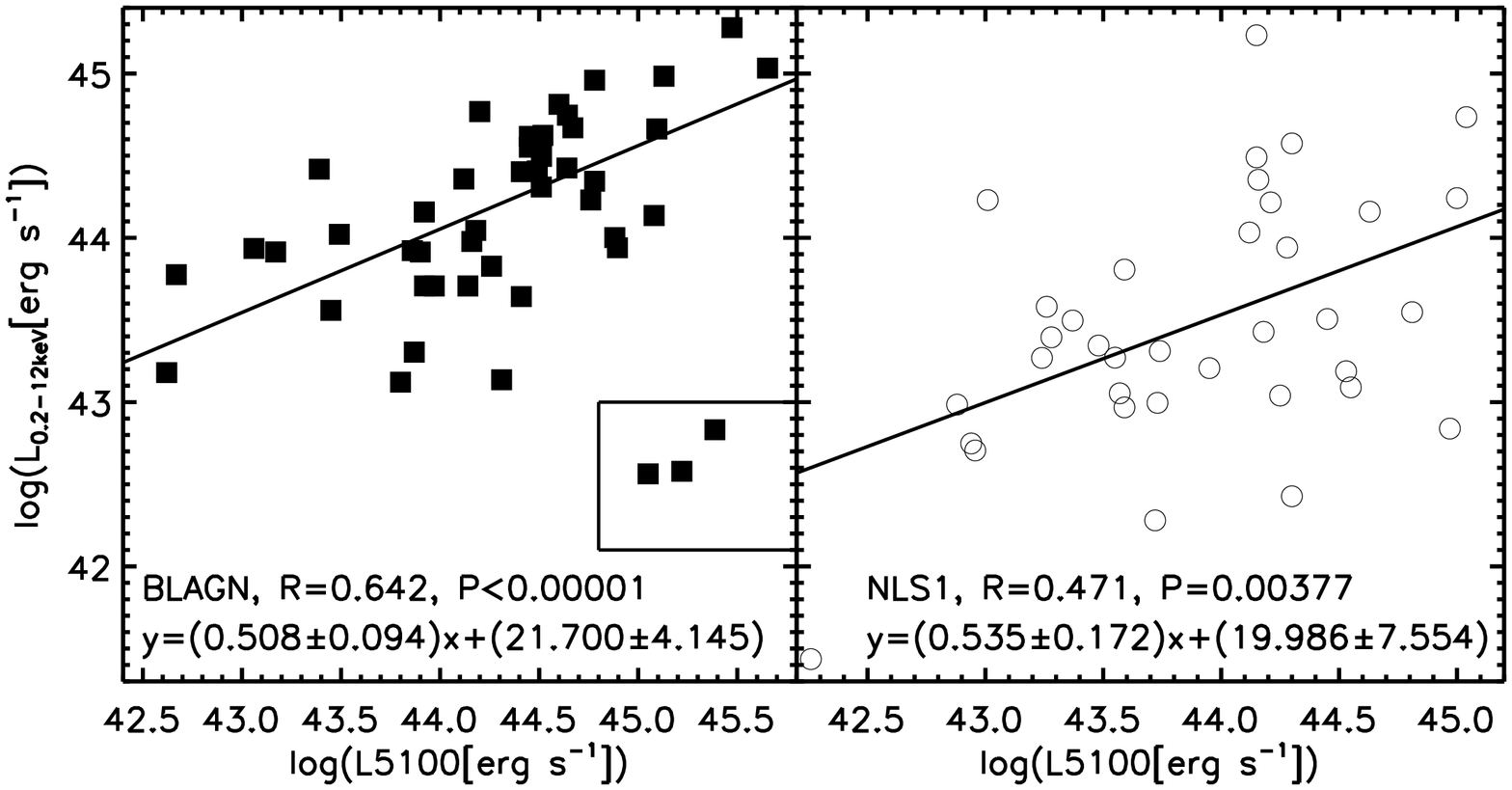}}
\caption{The luminosity comparison: luminosity at 6 $\mu$m, L6 vs. X-ray luminosity, L$_{\rm 0.2-12 keV}$ (upper panel), L6 vs. optical luminosity, L5100 (middle panel), and L$_{\rm 0.2-12 keV}$ vs. L5100 (bottom panel).  \label{fig:vp4}}
\end{figure*}

The AGN optical luminosity, L5100 and X-ray measured luminosity are compared with the AGN luminosity at 6 $\mu$m in Fig.~\ref{fig:vp4}. If we exclude the 3 outliers from the BLAGN plot (where log(L$\rm_{0.2-12 keV}$)<43 -- see the lines on the Fig.~\ref{fig:vp4}), then we obtain the X-ray--MIR correlation with Pearson correlation R=0.683, P<0.00001 for BLAGNs, while for NLS1 plot, we obtain trend with R=0.588, P=0.00016. 

The optical--MIR relation, obviously present in both NLS1 and BLAGN samples is shown in Fig.~\ref{fig:vp4} -- middle panel. 

The log(L5100)--log(L$\rm_{0.2-12 keV}$) relation is given in Fig.~\ref{fig:vp4} -- bottom panel. Again, if we exclude the three outliers for BLAGNs, where log(L$\rm_{0.2-12 keV}$)<43 (see the lines on the Fig.~\ref{fig:vp4}), we obtain the correlation R=0.642, P$<$0.00001, while for the NLS1 objects, the correlation is R=0.471, P=0.00377.

BLAGNs show somewhat higher correlations among X-rays, MIR and and optical luminosity than NLS1s (see the discussion in Sec.~\ref{sec:next}).

\begin{figure*} 
\centering
\rotatebox{90}{
\includegraphics[width=65mm]{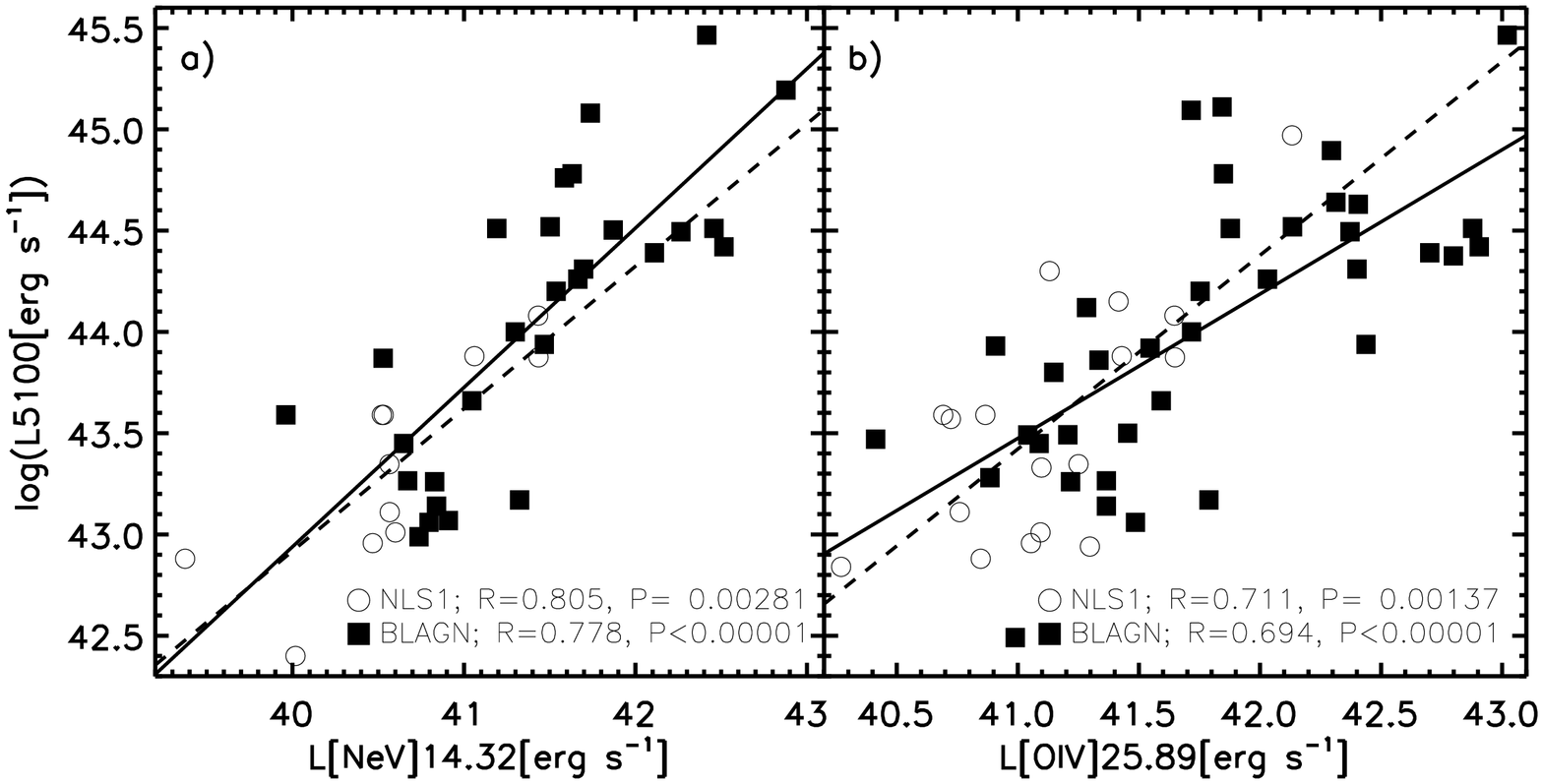}}
\rotatebox{90}{
\includegraphics[width=65mm]{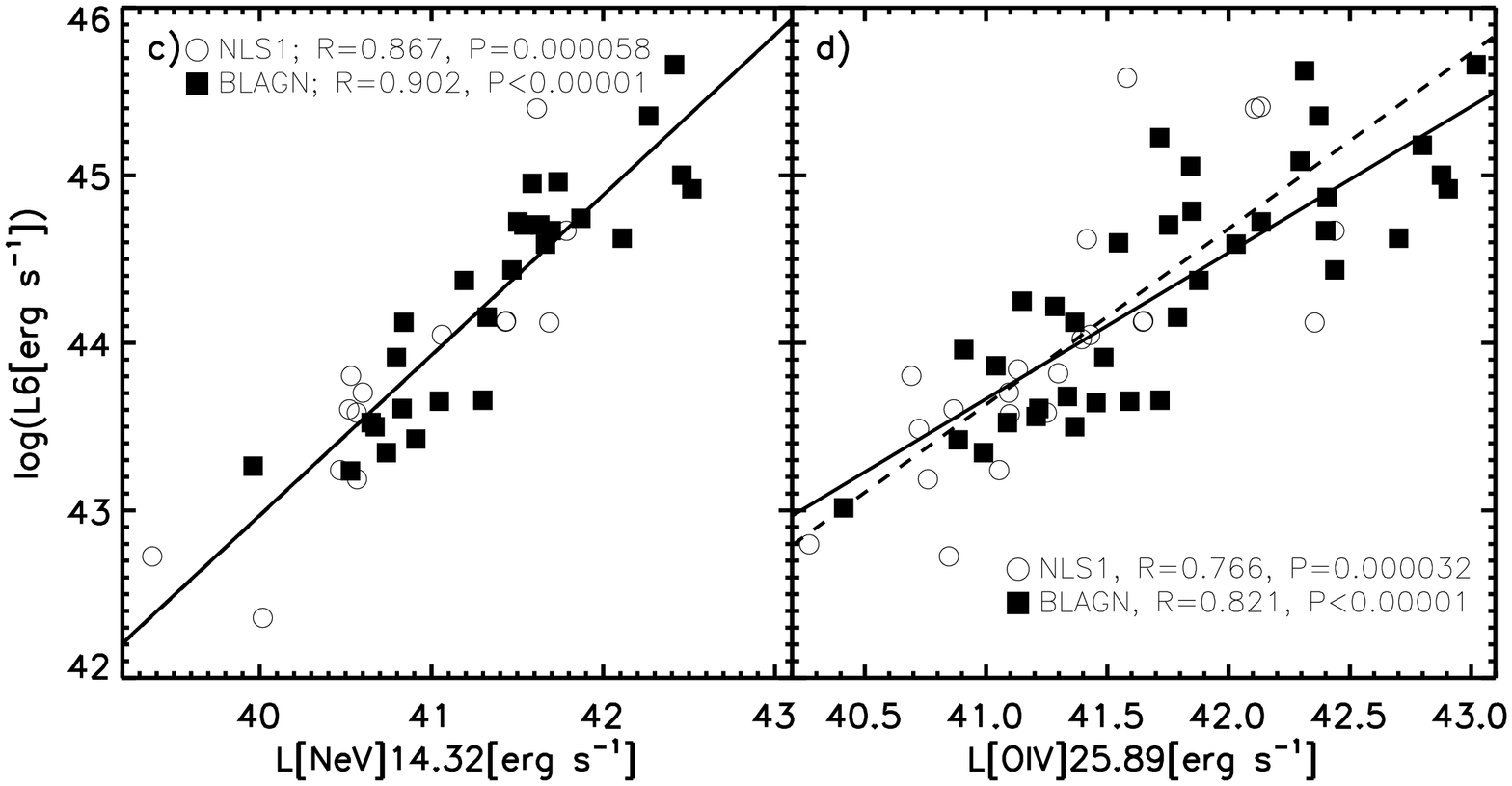}}
\caption{Relations: a) L5100--L[Ne\,V]14.32: NLS1: y=(0.703$\pm$0.173)x+ (14.794$\pm$7.022); BLAGN: y=(0.787$\pm$0.124)x+(11.472$\pm$5.169), b) L5100--L[O\,IV]25.89: NLS1: y=(0.956$\pm$0.244)x+ (4.207$\pm$10.045); BLAGN: y=(0.712$\pm$ 0.125)x+ (14.269$\pm$5.215), c) L6--L[Ne\,V]14.32: NLS1: y=(0.956$\pm$0.158)x+(4.725$\pm$6.462); BLAGN: y=(0.955$\pm$0.091)x+(4.753$\pm$3.777), and d) L6--L[O\,IV]25.89: NLS1: y=(1.049$\pm$0.196)x+(0.618$\pm$8.126); BLAGN: y=(0.872$\pm$0.103)x+(7.895$\pm$4.299). The dashed line is for NLS1, while full line is for BLAGNs. \label{fig:corlumx}}
\end{figure*}

Similarly as above, in Fig.~\ref{fig:corlumx} -- a) and b), we compare the relations between L5100 and the line luminosities: [Ne\,V] at 14.32 and [O\,IV] at 25.89 $\mu$m, for NLS1s and BLAGNs. The correlation is somewhat higher for [Ne\,V] than for [O\,IV] (see the caption and the legend). Also, in Fig.~\ref{fig:corlumx} -- c) and d) -- L6 is compared with the same lines. The correlations between coronal lines and L6 are stronger than with L5100. The correlations of L6 with coronal lines are higher for BLAGNs than for NLS1s, and again, higher for [Ne\,V] than for [O\,IV].

\begin{figure*}  
\centering
\rotatebox{90}{
\includegraphics[width=65mm]{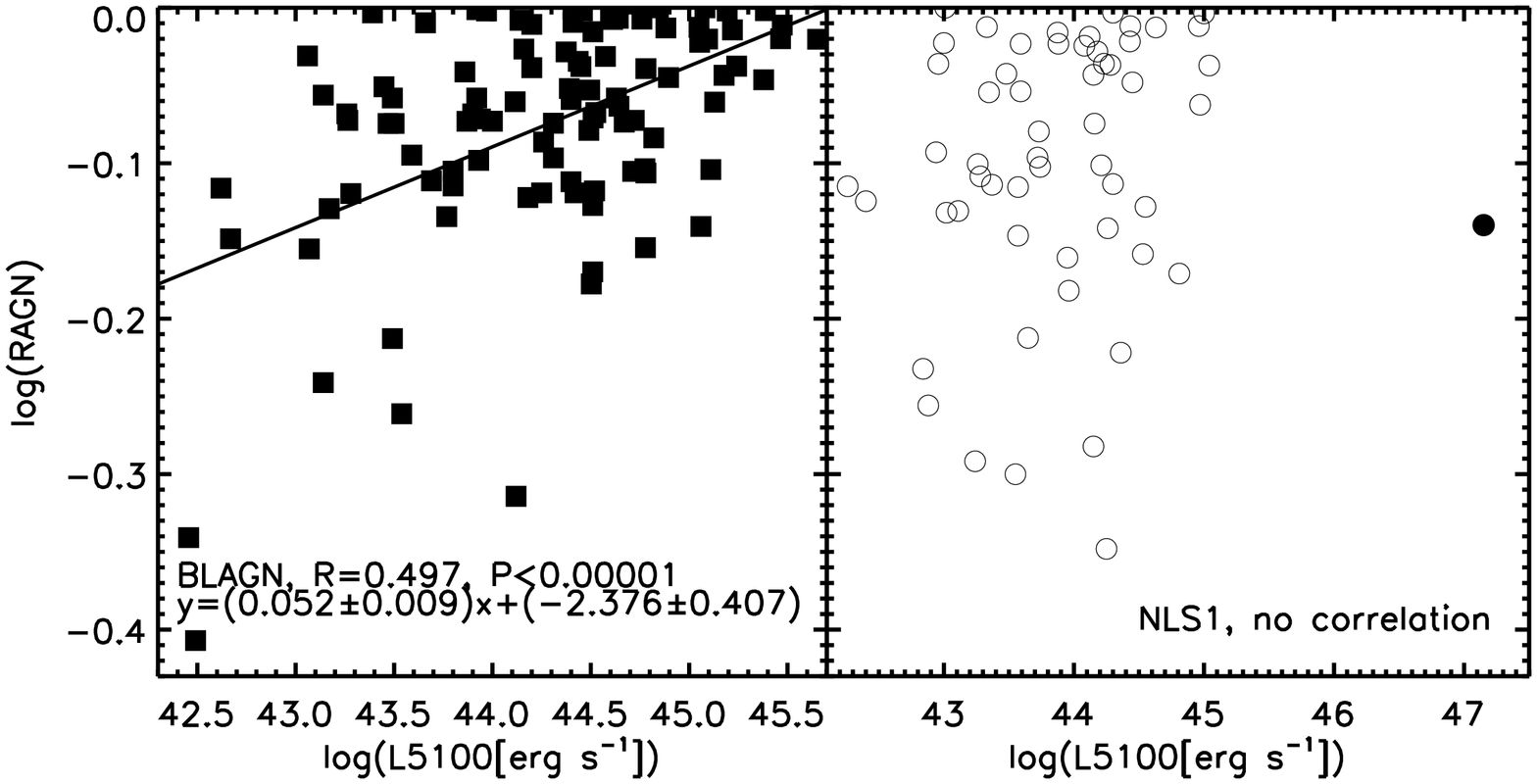}}
\rotatebox{90}{
\includegraphics[width=65mm]{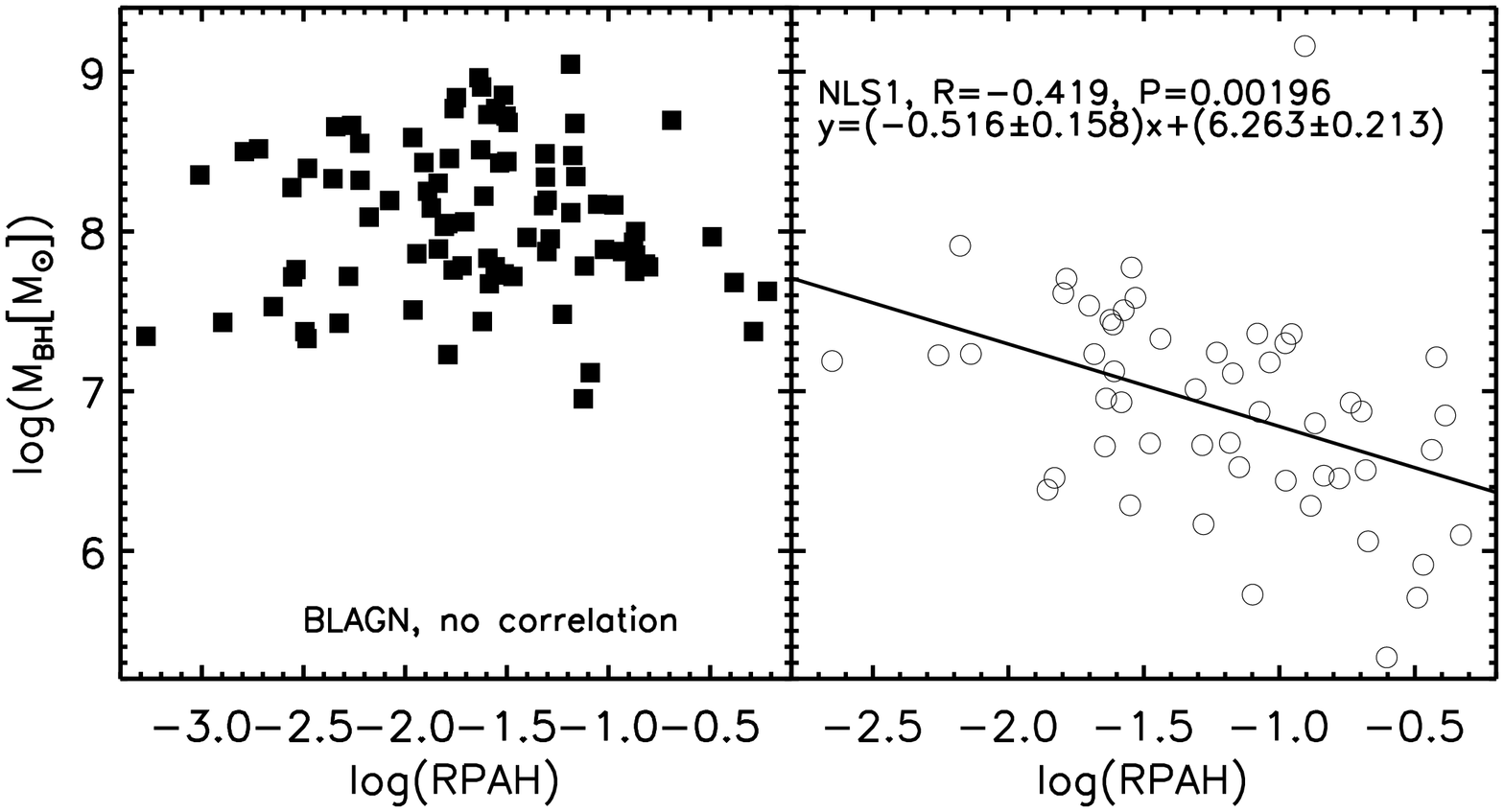}}
\caption{The comparison between optical and MIR parameters of NLS1s and BLAGNs: RAGN vs. L5100 (upper panel) and RPAH vs. M$_{\rm{BH}}$ (bottom panel). \label{fig:vp2}}
\end{figure*}

In Fig.~\ref{fig:vp2} -- upper panel, the plot RAGN--L5100 is given: a trend is present only for the BLAGNs, while for the NLS1s we do not notice any correlation. As well, when we exclude one data point (marked with black filled circle), where log(L5100)>47, from the NLS1s graph, we still do not find any trend for that plot. 

As we know the PAH luminosity grows with the AGN optical and MIR luminosity for both NLS1s and BLS1s \citep{Woo12}. In Fig.~\ref{fig:vp2} -- bottom panel -- we compare the well-known anti-correlation between RPAH and black hole mass \citep{Sani10,Lakicevic17}, for NLS1s (and we see the trend) and BLAGNs (where we do not notice any trend). That may indicate that NLS1s have PAHs destroyed by the AGN, as M$_{\rm{BH}}$ grows, while that is not the case for the BLAGNs. Possibly, the location of PAHs is further from the AGN in BLAGNs than in NLS1s.   

\begin{figure*}  
\centering
\rotatebox{90}{
\includegraphics[width=65mm]{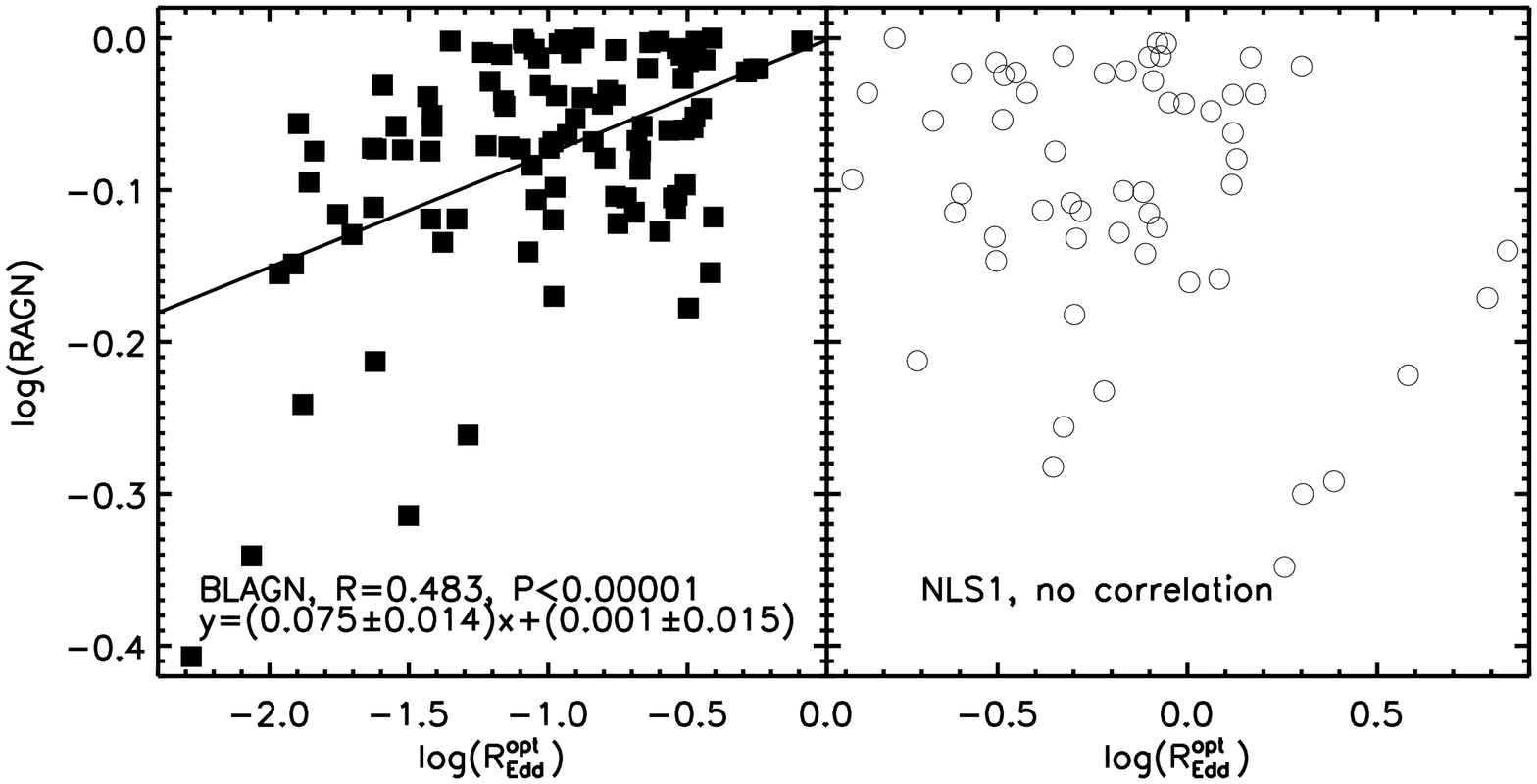}}
\rotatebox{90}{
\includegraphics[width=65mm]{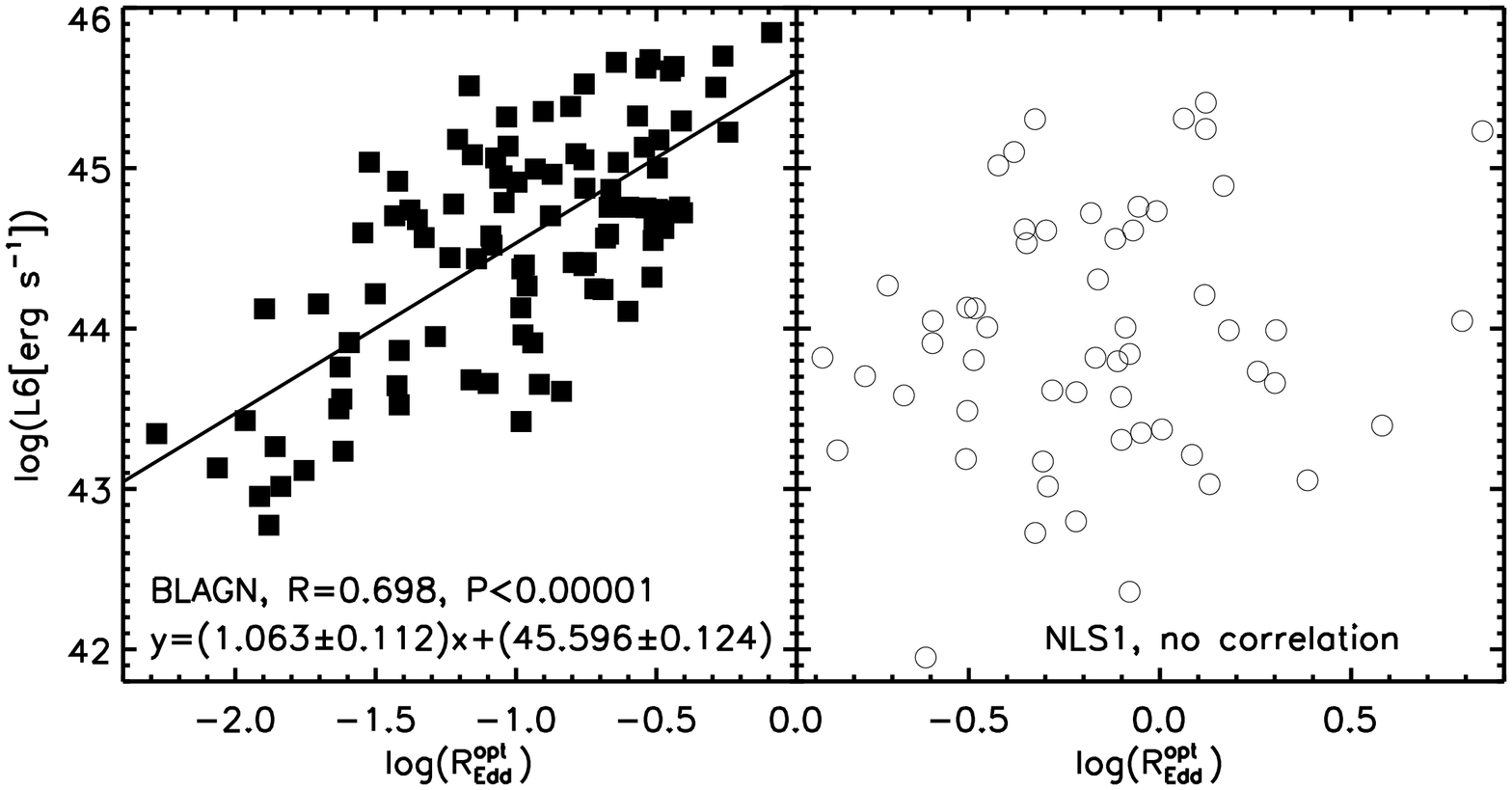}}
\caption{Top panel -- RAGN vs. optically derived Eddington ratio for BLAGNs and NLS1s, and bottom panel -- luminosity at 6 $\mu$m vs. optically derived Eddington ratio for BLAGNs and NLS1s.   \label{fig:vp3}}
\end{figure*}

We compare R$_{\rm Edd}$ with RAGN and with the luminosity at 6 $\mu$m in the Fig.~\ref{fig:vp3} (upper and bottom pannel, respectively). The trends are found only for BLAGNs, and not for NLS1s.

\subsubsection{Kinematic characteristics of the NLS1 and BLAGN sources} \label{sec:mirlines}
\begin{figure*}  
\centering
\rotatebox{90}{
\includegraphics[width=65mm]{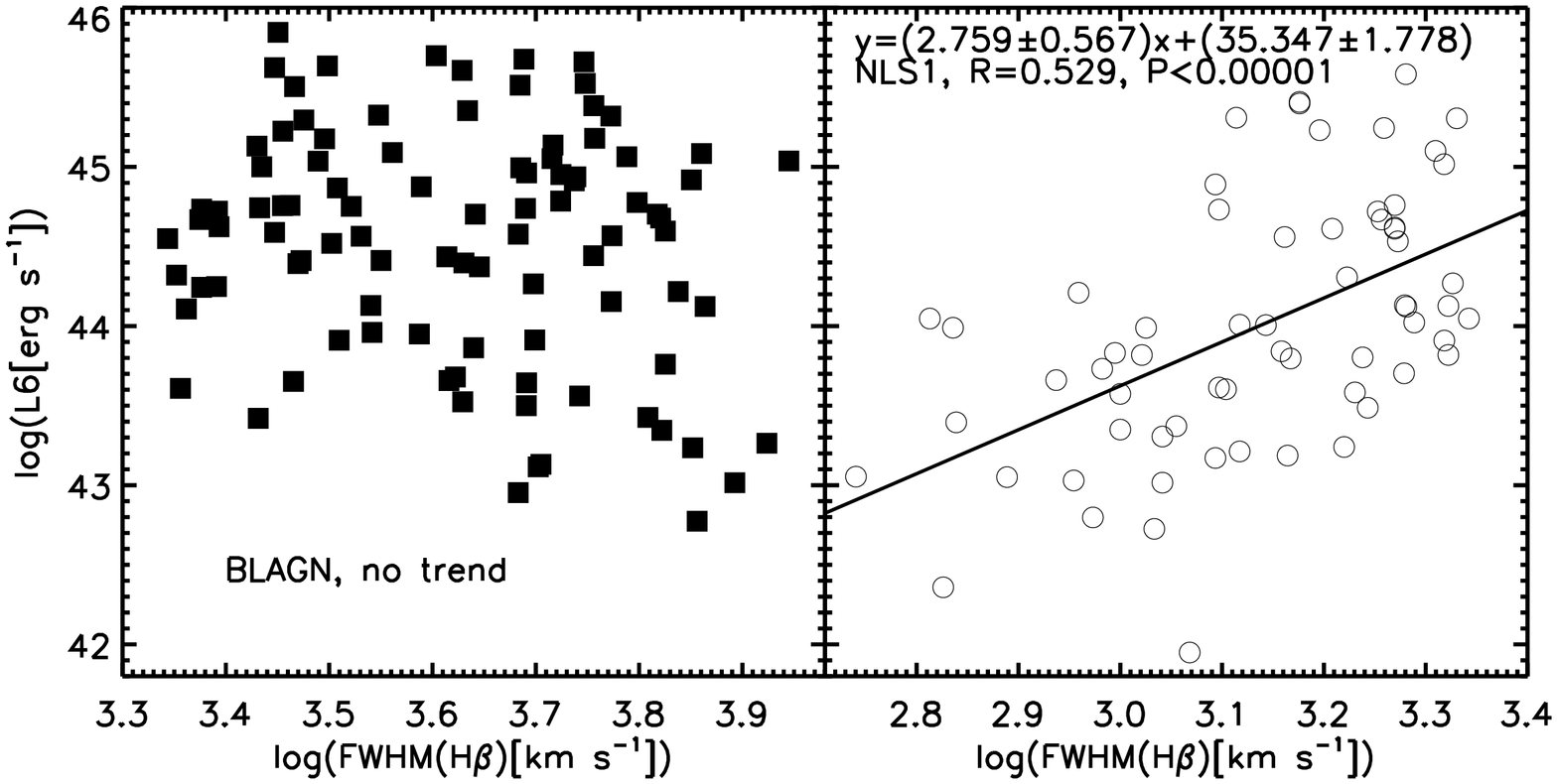}}
\rotatebox{90}{
\includegraphics[width=65mm]{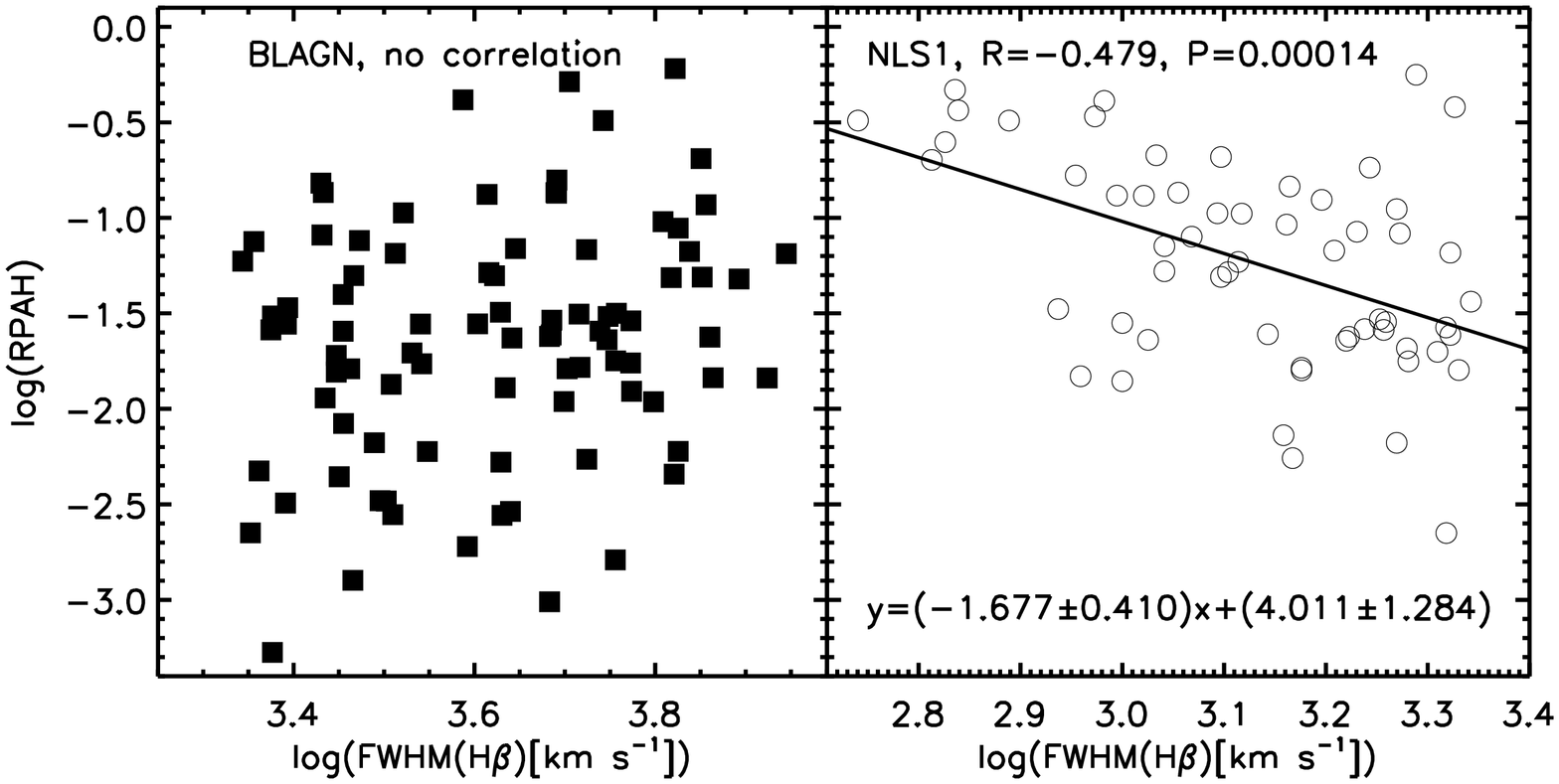}}
\caption{The upper panel -- FWHM(H$\beta$)--L6, for NLS1s and BLAGNs, and bottom panel -- FWHM(H$\beta$)--RPAH, for NLS1s and BLAGNs.   \label{fig:vp1}}
\end{figure*}

In order to explore difference between BLR kinematics in NLS1s and BLAGNs, we plot the FWHM(H$\beta$) as a function of the luminosity at 6 $\mu$m (see Fig.~\ref{fig:vp1}, upper panel). As it can be seen in Fig.~\ref{fig:vp1} (upper panel), there exists a correlation in the case of NLS1 sample, while there is no correlation in the BLAGN sample.

Also, we explore FWHM(H$\beta$) vs. RPAH (see Fig.~\ref{fig:vp1} -- bottom panel). We found an anti-correlation between RPAH and FWHM(H$\beta$) for NLS1, while there is no trend for the BLAGNs. Notice that \citet{Sani10} found that S1 objects with narrower FWHM(H$\beta$) have stronger PAH emission. That agrees with our plot for NLS1s and with the higher PAH abundance in NLS1 than in BLAGNs. Therefore, we conclude that this relation was not found in \citet{Lakicevic17} because the number of BLAGNs is 3 times higher than the number of NLS1s in that sample.   

\begin{figure*}  
\centering
\rotatebox{90}{
\includegraphics[width=65mm]{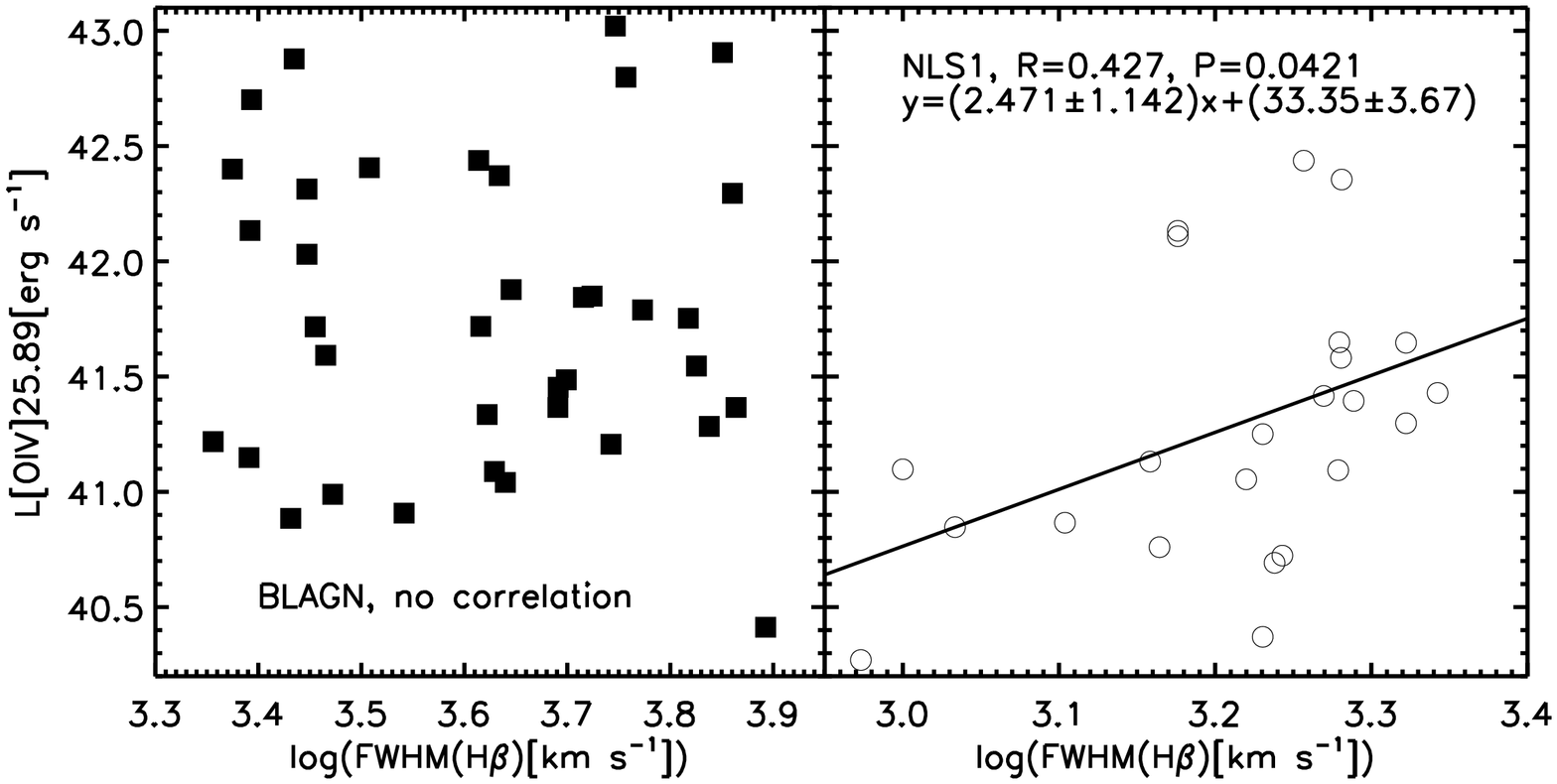}}
\rotatebox{90}{
\includegraphics[width=65mm]{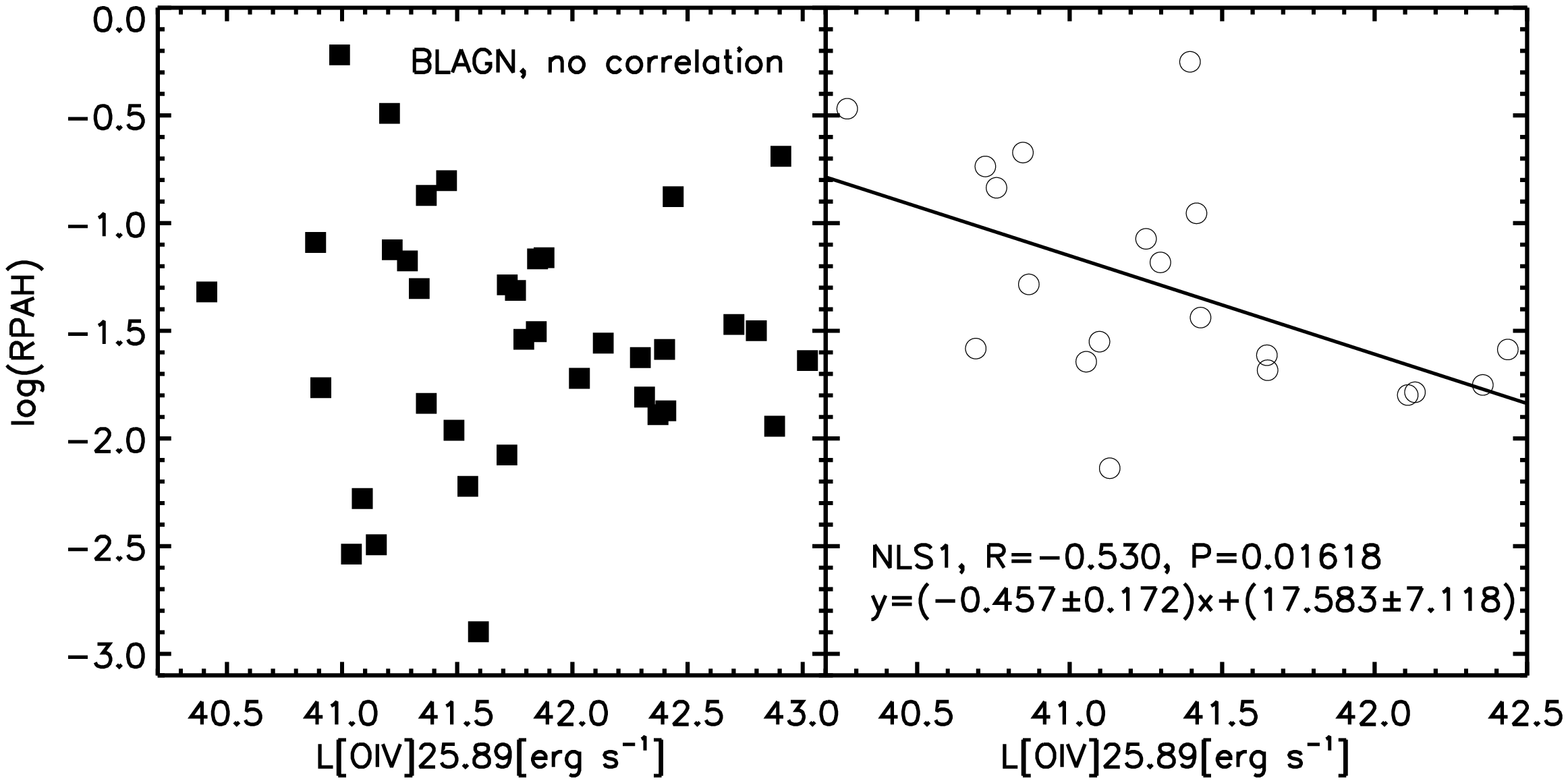}}
\caption{The luminosity of [O\,IV] line at 25.89 $\mu$m vs. FWHM(H$\beta$) (upper panel) and the luminosity of the [O\,IV] line at 25.89 $\mu$m vs. RPAH (bottom panel), for NLS1 and BLAGN objects.   \label{fig:coro1}}
\end{figure*}

\begin{figure*}  
\centering
\rotatebox{90}{
\includegraphics[width=65mm]{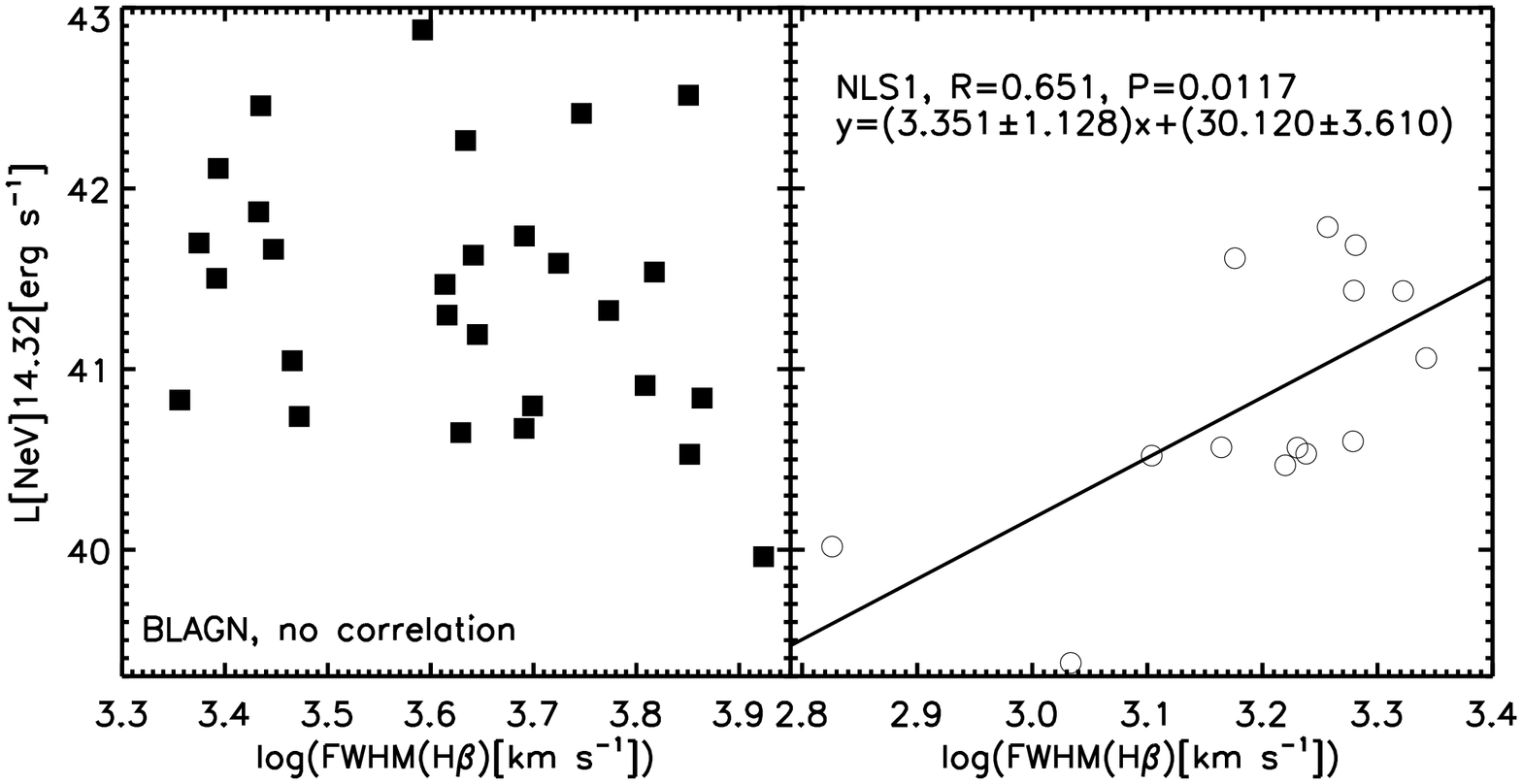}}
\rotatebox{90}{
\includegraphics[width=65mm]{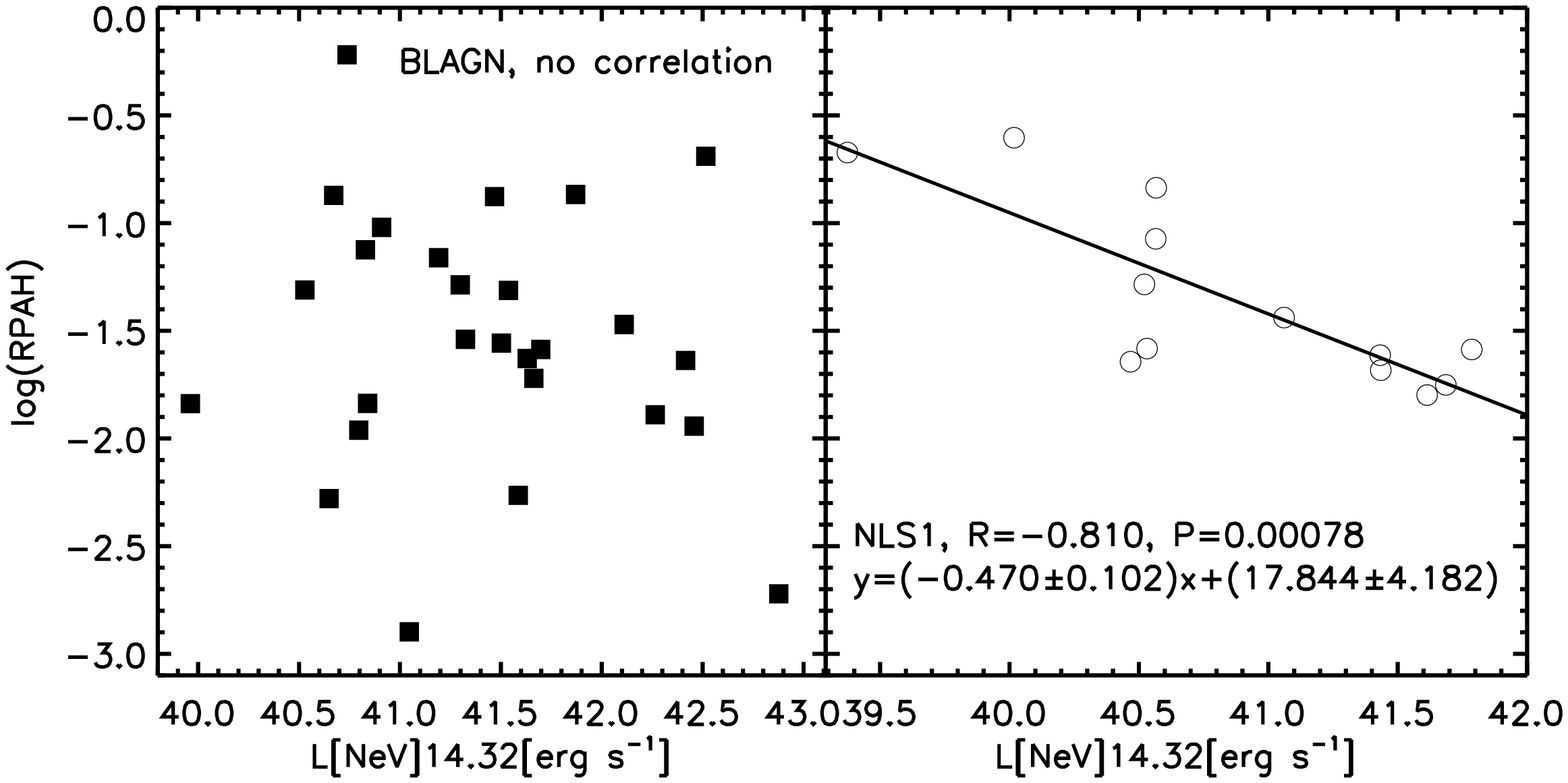}}
\caption{The luminosity of [Ne\,V] line at 14.32 $\mu$m vs. FWHM(H$\beta$) (upper panel) and the luminosity of [Ne\,V] line at 14.32 $\mu$m vs. RPAH (bottom panel), for NLS1 and BLAGN objects.   \label{fig:coro2}}
\end{figure*}

\begin{figure*}  
\centering
\rotatebox{90}{
\includegraphics[width=65mm]{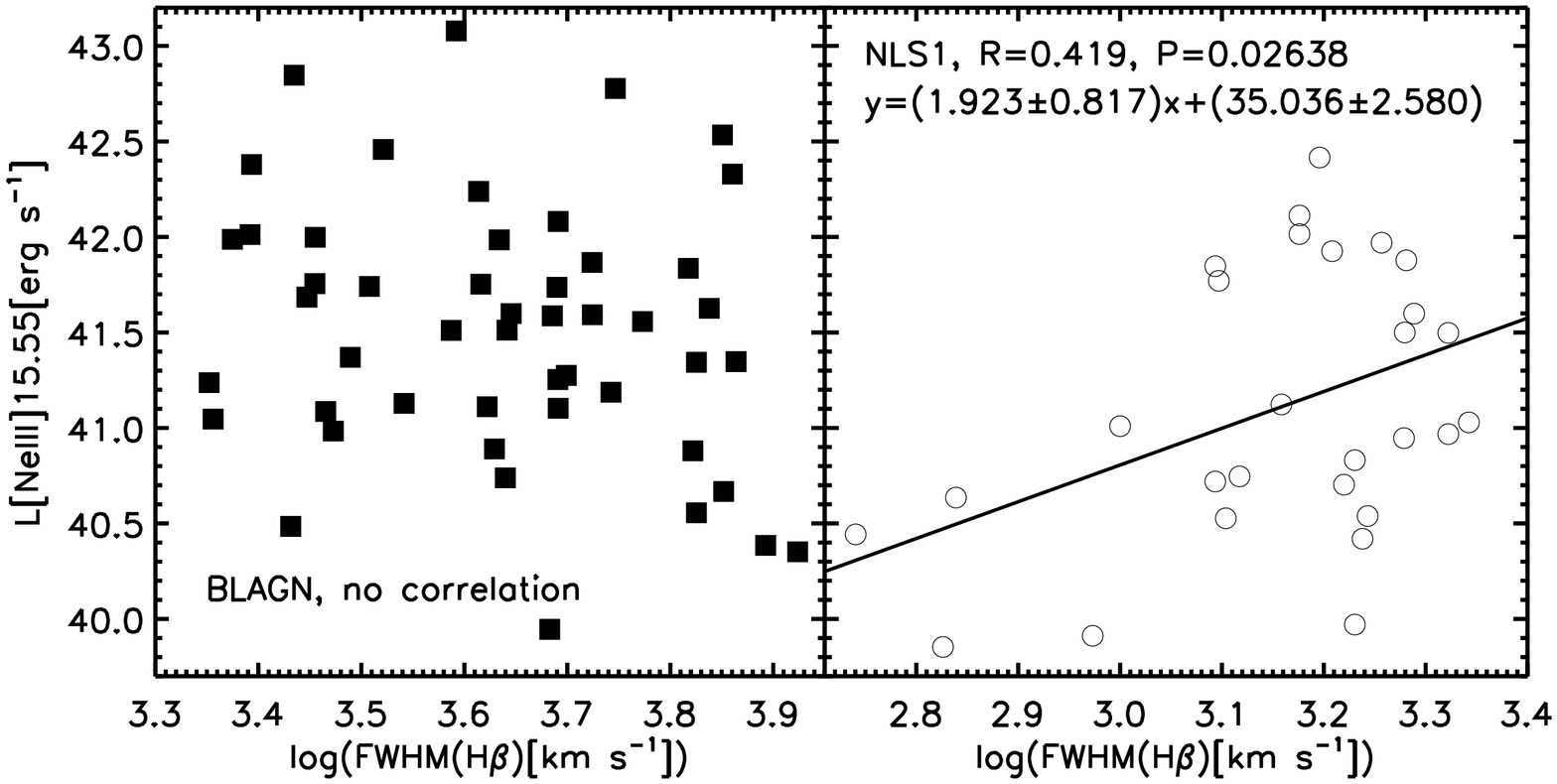}}
\rotatebox{90}{
\includegraphics[width=65mm]{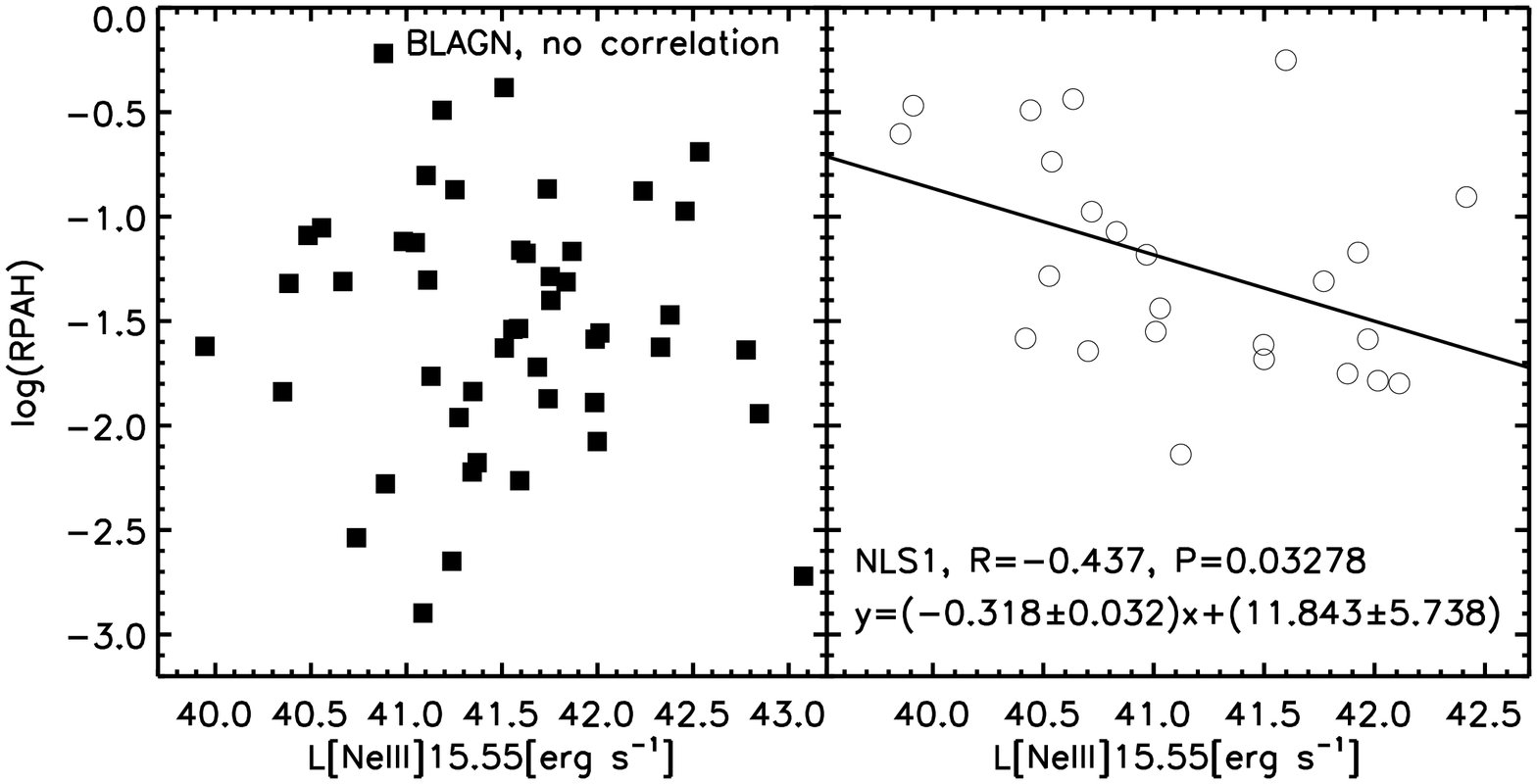}}
\caption{The luminosity of [Ne\,III] line at 15.55 $\mu$m vs. FWHM(H$\beta$) (upper panel) and the luminosity of [Ne\,V] line at 15.55 $\mu$m vs. RPAH (bottom panel), for NLS1 and BLAGN objects.   \label{fig:mirl}}
\end{figure*}

The luminosities of MIR lines: [Ne\,V] at 14.32 $\mu$m (97 eV), [O\,IV] at 25.89 $\mu$m (54 eV), as well as [Ne\,III] at 15.55 $\mu$m (41 eV) lines are found to behave differently for the NLS1 and BLAGN samples, in dependence with FWHM(H$\beta$) and RPAH (Figs.~\ref{fig:coro1}, \ref{fig:coro2} and \ref{fig:mirl}). It seems that for NLS1s there exist certain trend between line luminosities and FWHM(H$\beta$) (positive correlation) and between line luminosities and RPAH (an anti-correlation), while for BLAGNs, there is no any trend. Although the number of data points is low, the plots for all those three spectral lines suggest the same conclusions. 

\begin{figure*}  
\centering
\rotatebox{90}{
\includegraphics[width=65mm]{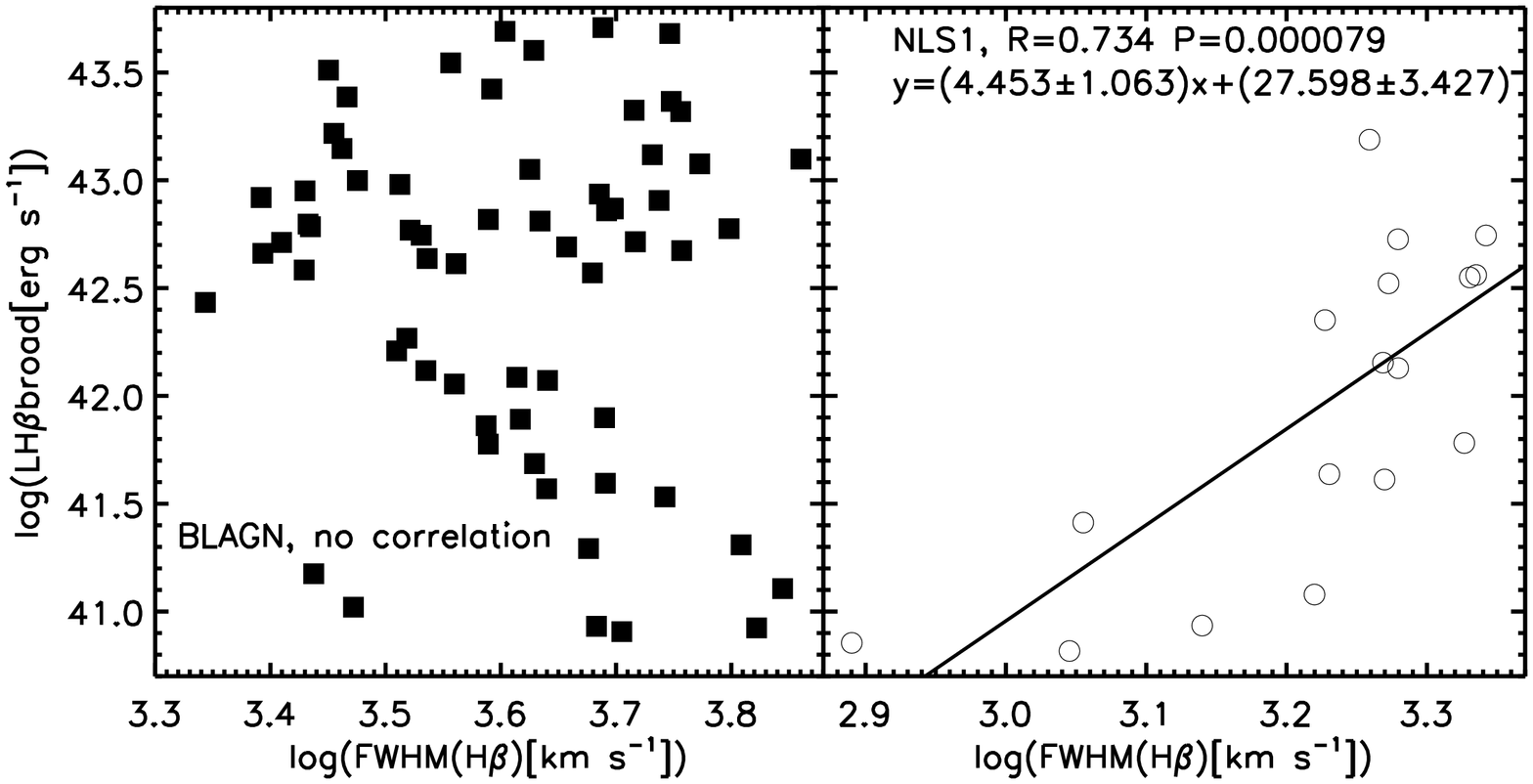}}
\rotatebox{90}{
\includegraphics[width=65mm]{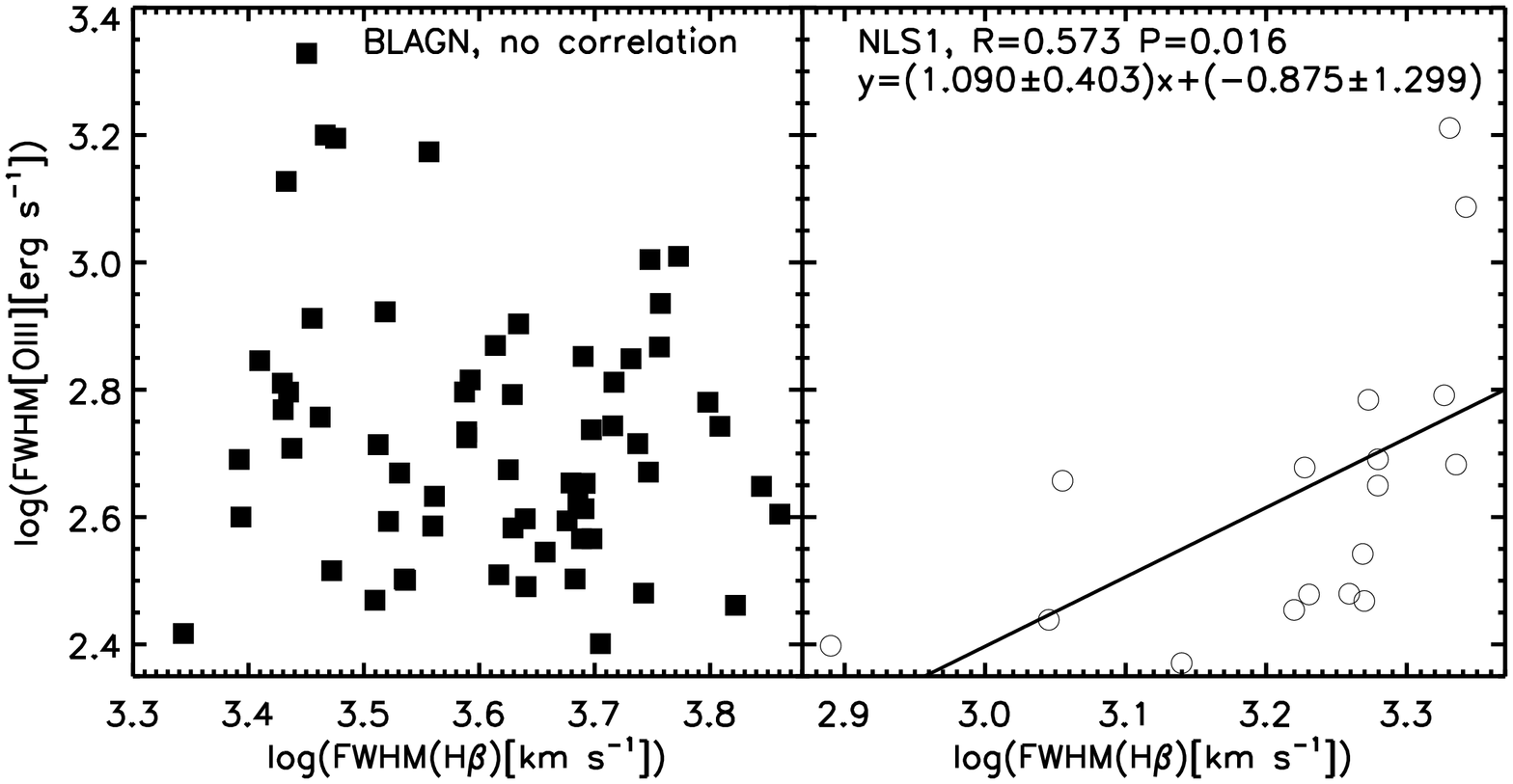}}
\rotatebox{90}{
\includegraphics[width=65mm]{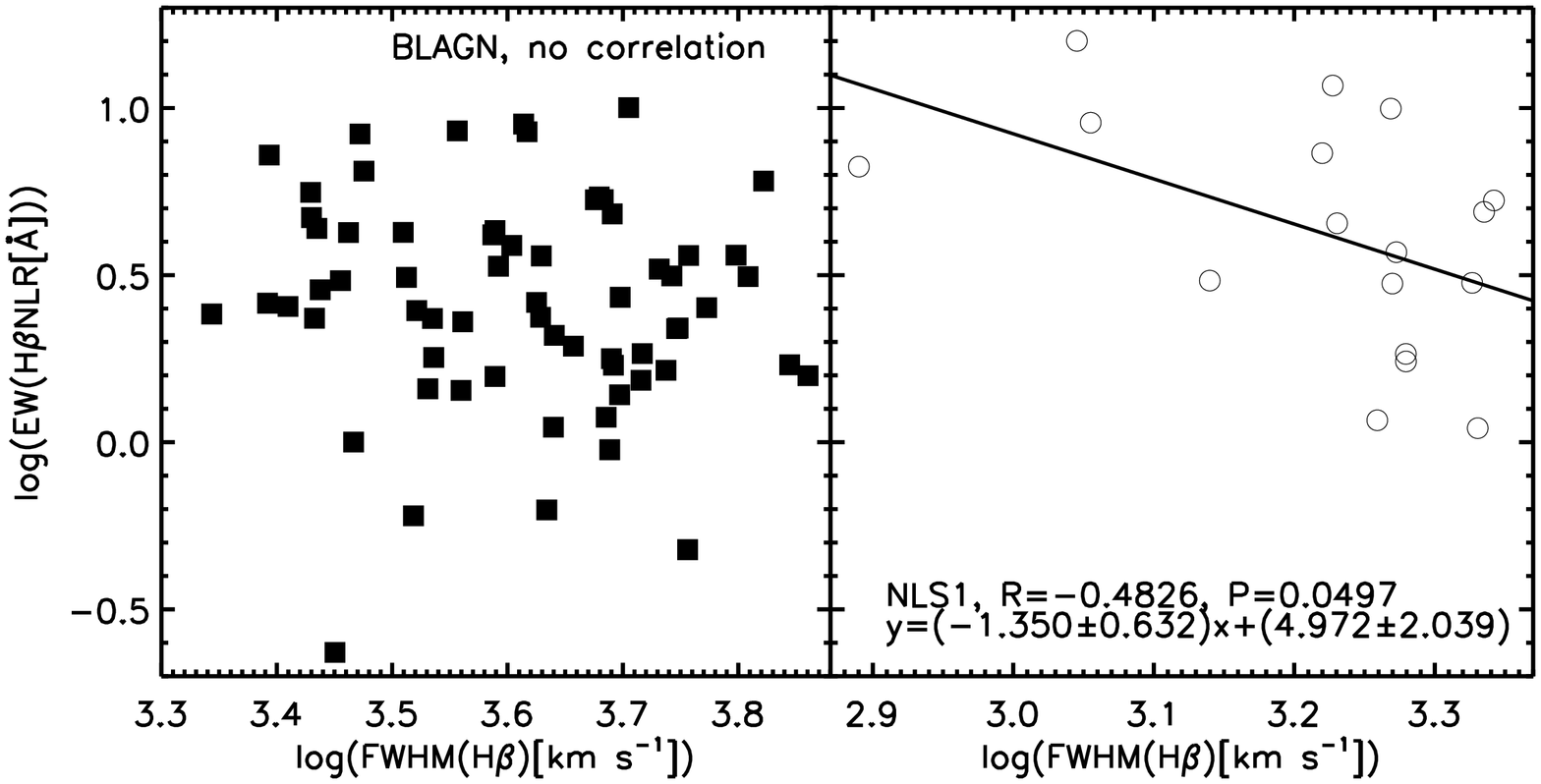}}
\caption{The evolution of the optical parameters: luminosity of H$\beta$ broad (upper panel), FWHM[O\,III] (middle) and EW(H$\beta$NLR) (bottom) of NLS1s with FWHM(H$\beta$). BLAGNs do not show any trends.   \label{fig:optikan}}
\end{figure*}

In Fig.~\ref{fig:optikan} we compared FWHM(H$\beta$) with some optical parameters for the objects from \citet{Lakicevic17}. The data for the upper panel are presented in Table~\ref{tab1sani10laki}, while the ones for the middle and bottom panels are taken from \citet{Lakicevic17}. It seems that, again, for NLS1s there exist positive trends: FWHM(H$\beta$)-luminosity of H$\beta$ broad (upper panel), FWHM(H$\beta$)-FWHM[O\,III] (middle) and an anti-correlation FWHM(H$\beta$)-EW(H$\beta$NLR)\footnote{Equivalent Width of H$\beta$NLR line -- EW(H$\beta$NLR).} (bottom). That, again, suggests a possibility of certain evolution of these parameters together with FWHM(H$\beta$) and M$\rm_{BH}$. 

\begin{figure*} 
\centering
\rotatebox{90}{
\includegraphics[width=65mm]{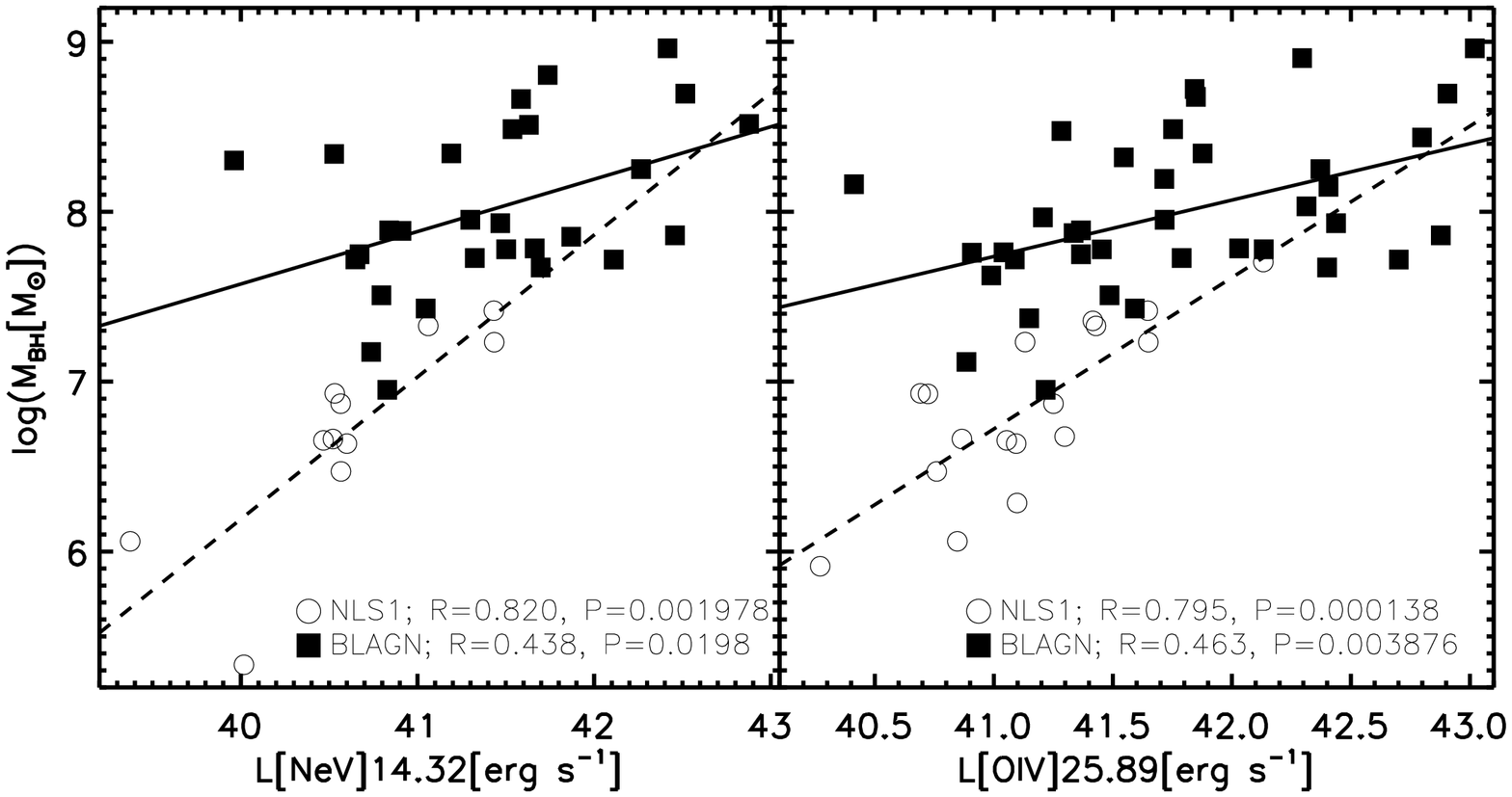}}
\caption{The relation between M$_{\rm BH}$ and luminosities of MIR lines for NLS1s and BLAGNs. The dashed line is for NLS1, while full line is for BLAGNs. \label{fig:masslum}}
\end{figure*}

\citet{Dasyra08} found that the luminosities of MIR coronal lines [Ne\,V] and [O\,IV] are correlated with the optically derived M$_{\rm BH}$s for 35 local AGNs. The Spearman rank coefficients were 0.76 and 0.69, respectively. In Fig.~\ref{fig:masslum}, we compare this relation for our two samples, for both [Ne\,V] and [O\,IV] lines, and obtain much stronger relation for NLS1s (R$\sim$0.8), while the relation for BLAGNs is weaker (R$\sim$0.4). Full lines are linear fits for BLAGN, while the dashed lines are for NLS1. Also, we notice that NLS1s have lower scatter about y values. 

\subsubsection{Additional optical parameters from the literature}
The dataset from \citet{Lakicevic17} contains only 17 NLS1s and 65 BLAGNs. The Man-Whitney U-test shows significant differences of log([O\,III]/H$\beta$NLR) among these two populations: P=0.0124. Mean and median log([O\,III]/H$\beta$NLR) for NLS1s are 0.684 and 0.59, while for BLAGNs they are: 0.95 and 1.03. That agrees with findings that this ratio is lower for the starburst dominated objects than for ''pure AGNs'' \citep{Popovic11}. However, we do not notice any differences in EW[O\,III] for these two populations, while for EW(H$\beta$NLR) the difference is quite high: P<0.00001: for NLS1s, mean is 5.52 and median is 4.52, while for BLAGNs, mean is 3.3, and the median is 2.6. 

\subsection{NLS1 and BLAGN correlations restricted by the redshift and L5100 ranges} \label{sec:novi}

As we are aware of the difference in redshifts in our NLS1 and BLAGN samples (see Fig.~\ref{fig:z}), we compared some BLAGN correlatons only in the NLS1 redshift range (z<0.39199). 87 out of 99 BLAGN objects have z<0.39199 (which is mutual NLS1 and BLAGN range). Using this new BLAGN range, we have not found the  M$_{\rm BH}$-RPAH trend. RPAH-FWHM(H$\beta$) was weakly related, but positively: R=0.282; P=0.013, on that interval. L6-FWHM(H$\beta$) was weakly, but negatively related: R=-0.224; P=0.037. RAGN-L5100 were related with R=0.477, P<0.00001. The optical accretion rate is correlated with L6 with R= 0.676; P<0.00001. Conclusion is that the redshift differences do not change much BLAGN correlations found in this sample.  

Similarly, as in previous paragraph, NLS1 and BLAGN mutual luminosity range is 42.455809$\le$log(L5100)$\le$45.652752 and contains 97 out of 99 BLAGNs. Again, M$_{\rm BH}$-RPAH trend was not found. RPAH-FWHM(H$\beta$) were weakly positively correlated R=0.218; P=0.043. L6-FWHM(H$\beta$) were not correlated. RAGN-L5100 had trend R=0.444; P<0.00001. Optical accretion rate was correlated with L6 with R=0.675; P<0.00001. Again, luminosity differences do not significantly change BLAGN trends. 

Finally, range 42.455809$\le$log(L5100)$\le$45.652752 contains 53 out of 56 NLS1 objects. M$_{\rm BH}$-RPAH trend for NLS1s was R=-0.524; P=0.00011. RPAH-FWHM(H$\beta$) correlation was: R=-0.481; P=0.00047. L6-FWHM(H$\beta$) trend was R=0.482, P=0.00025. RAGN-L5100 trend was not found. And the accretion rate was not related to L6 for objects in this sample. Like in the previous paragraph, luminosity difference does not change NLS1 correlations significantly.

\section{Discussion} \label{sec:discuss}

\subsection{The differences between NLS1s and BLS1s} \label{sec:prvoi}
In Sec.~\ref{sec:prvo} we showed the differences between NLS1s and BLS1s objects. One of the results is that the redshift is lower for the NLS1s than for the BLS1s. That phenomenon might be well understood in \citet{Rakshit17a}: on their Fig.~6 they show that the frequency of NLS1s remains relatively constant up to z<0.8, while the occurrence of BLS1s increases with the redshift. That may be the reason why we see that NLS1s have lower redshifts. However, the difference in z between NLS1s and BLS1s may also be the consequence of the selection effect. As NLS1s are less luminous than BLS1s (see Table~\ref{tab2}), we may not be able to observe NLS1s on the larger distances, while BLS1s could be observed. We do not consider the possibility that the redshift significantly influences the differences between NLS1s and BLS1s, since many studies showed that the NLS1s have remarkably different characteristics from BLS1s and because that redshift difference is not high. 

Our results show that most of the parameters are higher for BLS1s than for NLS1s (z, L5100, L6, RAGN, M$_{\rm BH}$, luminosity of [Ne\,V] and [O\,IV] coronal lines, X-ray luminosity (at the 0.2-12 keV band) and three hardness ratios), while the accretion rates R$_{\rm Edd}^{\rm opt}$, R$_{\rm Edd}^{\rm X}$ and RPAH are higher for the NLS1s. That agrees well with the results from the literature (Sec.~\ref{sec:intro}) and with the expectations of BLS1s being more massive and luminous objects with less starbursts than NLS1s. Although the NLS1s have lower RAGNs than BLS1s, the RAGNs for NLS1s are also usually quite high (upper panel of Fig.~\ref{fig:vp3}).   
 
\subsection{Comparison between X-ray, optical and MIR spectral characteristics of NLS1s and BLAGNs} \label{sec:next}
In Fig.~\ref{fig:vp4} -- the connection between the X-ray, optical and MIR luminosities are compared. If we exclude the 3 outliers from the upper and bottom BLAGNs panels, we notice that BLAGNs have higher correlations than NLS1s. Perhaps that could be explained by the higher strength of the BLAGN central sources, or that BLAGNs have more defined structures. 

Similarly, a relations between L5100 and the coronal line luminosities ([Ne\,V] at 14.32 and [O\,IV] at 25.89 $\mu$m) are found for both NLS1 and BLAGN samples (Fig.~\ref{fig:corlumx} -- a) and b) -- respectively). In Fig.~\ref{fig:corlumx} -- c) and d) -- L6 is compared with the same lines, respectively. L6 have stronger relations to coronal line luminosities than L5100. We can conclude that an AGN is the main heating source for both optical and MIR continuum. 

In Fig.~\ref{fig:vp2} -- top panel, a trend among RAGN and L5100 is presented: it exists for BLAGNs, but not for NLS1s. It might be that BLAGNs have better defined structure, or having AGN radiation more penetrating. 

As it can be seen in the Fig.~\ref{fig:vp2} -- bottom panel -- the M$_{\rm BH}$-RPAH relation is present only for the NLS1 objects. The PAHs destroyal is believed to exist close to AGNs \citep{ODowd09,Sales10,Diamond10}. Our results show that PAHs are significantly destroyed, while M$_{\rm BH}$ increases, for NLS1s. There might be a physical distance among PAHs (or starbursts) and the AGNs in BLAGN objects. Perhaps this anti-correlation may be explained by the fact that NLS1s have higher accretion rates and more spirals, therefore eating more PAH.  

In Fig.~\ref{fig:vp3} -- upper and bottom panels, we see that the optically derived accretion rate, R$_{\rm Edd}^{\rm opt}$, is somewhat dependent on the RAGN and L6, respectively, but only for the BLAGN objects, while NLS1s do not show that trend. Possible explanation for that phenomenon is connected with the recent study of \citet{Du18}, who showed that the BLR radius, R${\rm_{BLR}}$, is not related to the optical luminosity for the NLS1 objects (objects with high accretion rate). According to their relation (3), the R${\rm_{BLR}}$ is related with luminosity and with accretion rate. If R${\rm_{BLR}}$ does not depend on the luminosity, then the accretion rate might not be depending on the luminosity, as well, and the relation R${\rm_{BLR}}$-L5100 (Equation~\ref{eq:dr}) may not be correct for NLS1s. 

\subsubsection{Kinematic differences between NLS1s and BLAGNs} \label{kine}
In the Sec.~\ref{sec:drugo} there are given various correlations among NLS1 and BLAGN parameters. Firstly, Fig.~\ref{fig:vp1} -- upper panel presents a trend for NLS1s -- among FWHM(H$\beta$) and luminosity at 6 $\mu$m, L6. A somewhat similar comparison is done by \citet{Popovic11}, where the connection between FWHM(H$\beta$) and L5100 was found to exist in the AGNs where the starbursts are present (high correlation coefficient: R=0.74). They discussed this phenomenon and proposed the possibility that the starbursts influence both FWHM(H$\beta$) and L5100 in that group of AGNs. Here we suspect that the L6-FWHM(H$\beta$) and L5100-FWHM(H$\beta$) connection may have the different origin than the influence of the starburst: having in mind the relations among coronal line luminosity and FWHM(H$\beta$) (Figs.~\ref{fig:coro1} and ~\ref{fig:coro2}), coronal line luminosity--L5100 relation (Fig.~\ref{fig:corlumx} a) and b)), as well as coronal line luminosity--L6 relation (Fig.~\ref{fig:corlumx} c) and d)). We may doubt that the BH grow (not the starburst) is the source that increases luminosities and FWHM(H$\beta$) in NLS1 objects. AGN may be evolving with M$_{\rm BH}$, FWHM(H$\beta$) and luminosities grow, in NLS1s. Thus, L5100 and FWHM(H$\beta$) are related, but this L5100-FWHM(H$\beta$) relation is probably a secondary, not the primary relation. Perhaps that is also the case for the S1s with log([O\,III]5007/H$\beta$NLR)<0.5, that show L5100-FWHM(H$\beta$) correlation in \citet{Popovic11}. Curiously is that we have not found any correlation between FWHM(H$\beta$) and L5100 for our NLS1 neither BLAGN sample. 

Having in mind that the black hole mass depends on FWHM and luminosity as:
\begin{equation}
\rm M_{\rm BH} \sim L5100^{0.65} FWHM\left(H\beta\right)^{2} \label{eq:pr}
\end{equation}
\citep{Sani10}, we may consider why luminosity is correlated with FWHM(H$\beta$) in NLS1. The cause could be that the mass range is low and that FWHM(H$\beta$) is less dominant in NLS1 than in BLAGN, therefore FWHM(H$\beta$) is correlated to luminosity in NLS1s. Also, according to relation for the R${\rm_{BLR}}$ from \citet{Bentz09}: 
\begin{equation}
{\rm log \left(R_{BLR}\right)\Equ K} \Plus \alpha {\rm log\left(L5100\right)} \label{eq:dr},
\end{equation} where $\alpha \sim$0.6, and knowing that L5100 is higher for the BLAGN than in NLS1 (that we showed in Sec.~\ref{sec:prvo}) we may conclude that R${\rm_{BLR}}$ is larger for BLAGNs than NLS1s. Therefore, we may expect that together with the FWHM(H$\beta$) grow, R${\rm_{BLR}}$ also extend.

In the other hand, in the case of virialization (gravitationally driven motion), the Equations~\ref{eq:pr} and \ref{eq:dr} can lead us to FWHM(H$\beta$)$^{2} \sim$ ${\rm M_{\rm BH}}$ R${\rm_{BLR}^{-1.08}}$, which may be in contradiction with the previous assumption that R${\rm_{BLR}}$ grows with FWHM(H$\beta$). Again, the explanation might be that the Equation~\ref{eq:dr} does not work for the NLS1s \citep{Du18}. 

The trend FWHM(H$\beta$) -- RPAH, in Fig.~\ref{fig:vp1} -- bottom panel, only for NLS1 objects, also suggests that there is an interaction among starbursts and the AGNs, for NLS1s. It seems that the PAHs get destroyed and heated/excited by the AGN \cite[PAH luminosity grows with the AGN luminosity;][]{Sani10} as FWHM(H$\beta$) (and M$_{\rm BH}$) grow. 

As we saw in Figs.~\ref{fig:coro1},~\ref{fig:coro2} and \ref{fig:mirl}, the correlations among the MIR line luminosities of [Ne\,V], [O\,IV] and [Ne\,III] are present with FWHM(H$\beta$) and RPAH, but only for the NLS1s. This seems to agree with the evolution of FWHM(H$\beta$) with L6 and L5100 that we mentioned above. [Ne\,V] and [O\,IV] are coronal lines and can only be excited by the AGN. As M$_{\rm BH}$ grows, FWHM(H$\beta$) increases, the continuum and line luminosities increase, while RPAH drops, for NLS1 objects.   

In Fig.~\ref{fig:optikan} -- upper panel, a trend between FWHM(H$\beta$) and luminosity of H$\beta$ broad, seems to exist, for the NLS1s (R=0.734). That relation, for NLS1s, is already discussed by \citet{Veron01}, \citet{Zhou06} and the references within. \citet{Zhou06} found much weaker trend (R=0.29) and they supposed that this correlation is caused by the existence of upper limits in the accretion rate in Eddington units. \citet{Veron01} obtained higher correlation (R=0.76), both for NLS1s and BLAGNs. In their Fig.~15, one can see that BLAGNs in our range of log H$\beta$ luminosity (40.2-43.8), and even further, also do not have correlation LH$\beta$-FWHM(H$\beta$). However at the higher luminosities, that trend exists. It may be that NLS1s are in the phase of M$\rm_{BH}$ grow, thus -- everything changes with FWHM(H$\beta$) grow (MIR and optical line and continuum luminosities increase; see above, FWHM[O\,III] grows -- middle pannel of Fig.~\ref{fig:optikan}, and EW(H$\beta$NLR) drops -- Fig.~\ref{fig:optikan} -- bottom panel), while for BLAGNs -- that evolution does not exist. Could it all be explained only by the starburst presence in NLS1s? Some relations could be explained on that way, but the ones with coronal lines, FWHM(H$\beta$)--LH$\beta$ broad and FWHM[O\,III]--FWHM(H$\beta$) can not. [O\,III] is more likely connected with the AGN than with the starburst. 

BLR-NLR connection, found in NLS1 objects (FWHM[O\,III]--FWHM(H$\beta$), EW(H$\beta$NLR)--FWHM(H$\beta$), luminosity of MIR lines--FWHM(H$\beta$)), is also important to notice. FWHM[O\,III]--FWHM(H$\beta$) and luminosity of MIR lines--FWHM(H$\beta$) trends may exist because of FWHM is related to M$_{\rm BH}$. However, EW(H$\beta$NLR)--FWHM(H$\beta$) anti-correlation could be the consequence of AGN/starburst predominance, as H$\beta$NLR is related to the starbursts. 

The relations among the coronal line luminosities and M$_{\rm BH}$ are examined in the Fig.~\ref{fig:masslum}. According to these data, the trends are much higher for NLS1s than for the BLAGNs. It is doubtful if such connection can be used for the M$_{\rm BH}$ estimation for the BLAGNs. The reason for the better correlations for NLS1s may be that NLR (where the coronal lines come from) does not have the information about the BLR (that exists in BLAGNs). However, these trends agree with previous conclusions that the coronal line luminosity increase is connected with M$_{\rm BH}$ grow. 

\subsubsection{Summary of the differences between NLS1s and BLAGNs}
We confirmed and showed that BLS1s have higher optical, MIR and X-ray continuum luminosities, luminosities of MIR lines, M$_{\rm BH}$s, and RAGNs than NLS1s. NLS1s have higher accretion rates and RPAHs than BLS1s.  

We found several following relations that exist for NLS1s, but not for BLAGNs. Firstly, FWHM(H$\beta$) is correlated with line and continuum luminosities. Secondly, RPAH is anti-correlated with the line and continuum luminosities. Thirdly, FWHM(H$\beta$) is correlated to log (LH$\beta$ broad) and FWHM[O\,III] and anti-correlated with EW(H$\beta$NLR). One possible explanation for these relations for NLS1s is that these objects have their BHs growing together with all luminosities and FWHM(H$\beta$), while RPAHs drop. 

The relations among the MIR coronal line luminosities and M$_{\rm BH}$s are found to be stronger for the NLS1s than for the BLAGNs. It may be because NLR (where the coronal lines come from) does not have the information about the BLR (that exists in BLAGNs), therefore this method might not be suitable for the M$_{\rm BH}$ estimation for BLAGN objects.
 
\section{Conclusions} \label{sec:conclussion}
Here we investigated the differences in the optical, X-ray and MIR spectral characteristics between NLS1 and BLS1 and/or BLAGN objects. We found several relations which seem to exist for NLS1s, but not for BLAGNs: as FWHM(H$\beta$) grows, luminosities of optical and MIR continuum, luminosity of MIR lines, luminosity of H$\beta$ broad and FWHM[O\,III]-- grow as well, while RPAH and EW(H$\beta$NLR) drop. The reason for these trends in NLS1s is probably the black hole grow and consequently the luminosities and the FWHM(H$\beta$) grow, while BLAGNs do not show these trends because they have a different evolution and also possibly a different physical characteristics (for example starbursts far away from the AGN). 

Here we list some conclusions:
\begin{enumerate} 
\item Parameters: L5100, M$_{\rm BH}$, RAGN, L6, and luminosities of coronal lines [Ne\,V] and [O\,IV] are higher for the BLS1s than for the NLS1 objects. RPAH and accretion rates are higher for NLS1s than for BLS1s. Most of conclusions for BLS1s from this work most likely apply for the BLAGNs, since BLAGNs are brighter and more massive than BLS1s. 

\item BLS1s are harder sources than NLS1s. X-ray luminosity, L$_{\rm 0.2-12 keV}$ is higher for BLS1s than for NLS1s.

\item In NLS1s: the M$_{\rm BH}$ grow may be the reason for the FWHM(H$\beta$) and L5100 increase, therefore -- a reason for the correlation FWHM(H$\beta$)-L5100 \citep{Popovic11} and FWHM(H$\beta$)-L6 trend (this work). For the same reason, FWHM(H$\beta$) is also in trend with luminosity of H$\beta$ broad, the luminosity of MIR lines and FWHM[O\,III], while anti-correlated with EW(H$\beta$NLR) and RPAH.  

\item The anti-correlations among FWHM(H$\beta$) and coronal line luminosities with RPAH, for NLS1s suggest that AGN may destroy PAHs for these objects, as BH grows. 

\item Relations FWHM(H$\beta$)--FWHM[O\,III], FWHM(H$\beta$)--EW(H$\beta$NLR) and FWHM(H$\beta$)--luminosity of MIR lines -- show the connection between NLR and BLR in NLS1 objects.  

\item NLS1s have relation, M$_{\rm BH}$-RPAH (Pearson coefficient: P=-0.419), while BLAGNs do not show that trend. 

\item NLS1s have significantly stronger correlation between M$_{\rm BH}$ and the luminosity of coronal lines (R$\sim$0.8), than BLAGNs (R$\sim$0.4). That relation is proposed for the estimation of the M$_{\rm BH}$ \citep{Dasyra08}. Perhaps NLS1s have stronger trend because NLR (where the coronal MIR lines come from) does not contain the information about the BLR (that exist in BLAGN objects), while NLS1s might not have BLR. 

\end{enumerate}

\section*{Acknowledgments}
This work is part of the project (176001) "Astrophysical Spectroscopy of Extragalactic Objects" supported by the Ministry of Science of Serbia.

The Cornell Atlas of Spitzer/IRS Sources (CASSIS) is a product of the Infrared Science Center at Cornell University, supported by NASA and JPL.

Much of the analysis presented in this work was done with TOPCAT (\url{http://www.star.bris.ac.uk/\~mbt/topcat/}), developed by M. Taylor.

We thank dr Palle M{\o}ller for helpful advices. We thank the referee for careful reading our manuscript and constructive suggestions and comments. 

\appendix{

\section{[OIII]$\lambda$5007/H$\beta$ ratio for the remaining objects} \label{appen}
 
In this work we used 7 objects from \citet{Lakicevic17} that were not previously classified as NLS1 objects, although they have FWHM(H$\beta$)$\le$2200 km$^{-1}$. For that, it was needed to check if their ratio of fluxes of total [OIII]$\lambda$5007 and total H$\beta$ (where total applies to the sum of both broad and narrow components) is <3. In the Table~\ref{tabapp} we give the fluxes of these lines and their ratios. We concluded that these 7 objects are NLS1s. 
 
\begin{table*}
\caption{The fluxes of total H$\beta$, total [OIII]$\lambda$5007 and their ratios, for 7 remaining objects (see the text). These ratios are <3, their FWHM(H$\beta$)$\le$2200 km s$^{-1}$, therefore these objects can be classified as NLS1s.  \label{tabapp}}
\centering{
\begin{tabular}{|l|c|c|c|c|}
\hline\hline
NAME   &                     Plate-MJD-fiber&  FluxH$\beta^{\rm tot}$ &Flux[OIII]$\lambda$5007 &Flux[OIII]$\lambda$5007/FluxH$\beta^{\rm tot}$ \\ 
       &                                    & [10$^{-17}$ erg s$^{-1}$ cm$^{-2}$]  &[10$^{-17}$ erg s$^{-1}$ cm$^{-2}$] &    \\                   
\hline
  UM\_614                   &0530-52026-0165& 5475.64&      11813.1   &  2.157       \\ 
  MRK\_0707                 &0476-52314-0523& 7668.34&      3724.57   &  0.486        \\ 
  FBQS\_J1448+3559          &1383-53116-0476& 4153.63&      1588.97   &  0.382        \\ 
  FBQS\_J125807.4+232921    &2662-54505-0293& 1749.73&      174.46    &  0.10         \\ 
  SDSS\_J171207.44+584754.4 &0355-51788-0408& 1536.38&      482.91    &  0.314         \\
  {[HB89]}\_2233+134          &0739-52520-0388& 4479.16&      1182.3    &  0.264        \\ 
  SDSS\_J021446.99-003250.6 &4236-55479-0268& 157.024&      100.28    &  0.639         \\
 \hline
\end{tabular}}
\\
\smallskip
\end{table*}

}

\end{document}